\documentclass[traditabstract,epsf]{aa}

\def\m2s2{\hbox{\,m$^{2}$\,s$^{-2}$}} 

\PassOptionsToPackage{hyphens}{url}\usepackage[colorlinks = true, breaklinks]{hyperref}
\hypersetup{urlcolor=black,linkcolor=black,citecolor=black,colorlinks=true}
\usepackage{breakurl}

\usepackage{graphicx}
\usepackage{txfonts}
\usepackage{natbib}
\usepackage{verbatim}

\begin{document}
   \title{Transiting planets from WASP-South, Euler and TRAPPIST:}

   \subtitle{WASP-68\,b, WASP-73\,b  and WASP-88\,b, \\three hot Jupiters transiting evolved solar-type stars\thanks{The photometric time-series used in this work are only available at the CDS via anonymous ftp to \url{cdsarc.u-strasbg.fr (130.79.128.5)} or via \url{http://cdsarc.u-strasbg.fr/viz-bin/qcat?J/A+A/}}}

\author{
 	  L.~Delrez\inst{1},
	  V.~Van Grootel\inst{1}\thanks{Charg\'ee de recherches, Fonds de la Recherche Scientifique, FNRS, rue d'Egmont 5, B-1000 Bruxelles, Belgium},
	  D. R.~Anderson\inst{2},
	  A.~Collier-Cameron\inst{3},
	  A. P.~Doyle\inst{2},
	  A.~Fumel\inst{1},
          M.~Gillon\inst{1},
          C.~Hellier\inst{2},
          E.~Jehin\inst{1},
          M.~Lendl\inst{4},
          M.~Neveu-VanMalle\inst{4,6},
          P. F. L.~Maxted\inst{2},
          F.~Pepe\inst{4},
          D.~Pollacco\inst{5},
          D.~Queloz\inst{6,4},
          D.~S\'{e}gransan\inst{4},
          B.~Smalley\inst{2},
          A. M. S.~Smith\inst{7,2},
          J.~Southworth\inst{2},
          A. H. M. J.~Triaud\inst{8,4}\thanks{Fellow of the Swiss National Science Foundation.},
          S.~Udry\inst{4},
          R. G.~West\inst{5}
          }

\offprints{ldelrez@ulg.ac.be}
\institute{ 
       	    $^1$ Institut d'Astrophysique et G\'eophysique, Universit\'{e} de Li\`{e}ge, all\'{e}e du 6 Ao\^{u}t 17, B-4000 Li\`{e}ge, Belgium\\
	    $^2$ Astrophysics Group, Keele University, Staffordshire, ST5 5BG, UK\\
	    $^3$ SUPA, School of Physics and Astronomy, University of St. Andrews, North Haugh, Fife, KY16 9SS, UK\\
	    $^4$ Observatoire de Gen\`eve, Universit\'{e} de Gen\`{e}ve, 51 Chemin des Maillettes, 1290 Sauverny, Switzerland\\
	    $^5$ Department of Physics, University of Warwick, Coventry CV4 7AL, UK\\
	    $^6$ Cavendish Laboratory, Department of Physics, University of Cambridge, JJ Thomson Avenue, Cambridge, CB3 0HE, UK\\
	    $^7$ N. Copernicus Astronomical Centre, Polish Academy of Sciences, Bartycka 18, 00-716 Warsaw, Poland\\
	    $^8$ Department of Physics and Kavli Institute for Astrophysics \& Space Research, Massachusetts Institute of Technology, Cambridge, MA 02139, USA\\	     
	    }
\date{Received date / accepted date}
\authorrunning{L. Delrez et al.}
\titlerunning{WASP-68\,b, WASP-73\,b  and WASP-88\,b}

  \abstract
  {We report the discovery by the WASP transit survey of three new hot Jupiters, WASP-68 b, WASP-73 b and WASP-88 b. \hbox{WASP-68 b} has a mass of 0.95 $\pm$ 0.03 $M_{\mathrm{Jup}}$, a radius of \hbox{$1.24_{-0.06}^{+0.10}$ $R_{\mathrm{Jup}}$}, and orbits a V=10.7 G0-type star (\hbox{1.24 $\pm$ 0.03 $M_{\odot}$}, \hbox{$1.69_{-0.06}^{+0.11}$ $R_{\odot}$}, \hbox{$T_{\mathrm{eff}}$ = 5911 $\pm$ 60 K}) with a period of \hbox{5.084298 $\pm$ 0.000015 days}. Its size is typical of hot Jupiters with similar masses. \hbox{WASP-73 b} is significantly more massive (\hbox{$1.88_{-0.06}^{+0.07}$ $M_{\mathrm{Jup}}$}) and slightly larger ($1.16_{-0.08}^{+0.12}$ $R_{\mathrm{Jup}}$) than Jupiter. It orbits a V=10.5 F9-type star (\hbox{$1.34_{-0.04}^{+0.05}$ $M_{\odot}$}, $2.07_{-0.08}^{+0.19}$ $R_{\odot}$, \hbox{$T_{\mathrm{eff}}$ = 6036 $\pm$ 120 K}) every 4.08722 $\pm$ 0.00022 days. Despite its high irradiation ($\sim$2.3 $10^{9}$ erg $\mathrm{s}^{-1} \mathrm{cm}^{-2}$), WASP-73 b has a high mean density ($1.20_{-0.30}^{+0.26}$ $\rho_{\mathrm{Jup}}$) that suggests an enrichment of the planet in heavy elements. \hbox{WASP-88 b} is a \hbox{0.56 $\pm$ 0.08 $M_{\mathrm{Jup}}$} planet orbiting a V=11.4 F6-type star (1.45 $\pm$ 0.05 $M_{\odot}$, $2.08_{-0.06}^{+0.12}$ $R_{\odot}$, $T_{\mathrm{eff}}$ = 6431 $\pm$ 130 K) with a period of \hbox{4.954000 $\pm$ 0.000019 days}. With a radius of $1.70_{-0.07}^{+0.13}$ $R_{\mathrm{Jup}}$, it joins the handful of planets with super-inflated radii. The ranges of ages we determine through stellar evolution modeling are 4.2-8.3 Gyr for WASP-68, 2.7-6.4 Gyr for WASP-73 and 1.8-5.3 Gyr for WASP-88. WASP-73 appears to be a significantly evolved star, close to or already in the subgiant phase. WASP-68 and WASP-88 are less evolved, although in an advanced stage of core H-burning.
  }

   \keywords{planetary systems --
                stars: individual: WASP-68 --
                stars: individual: WASP-73 --
                stars: individual: WASP-88 --
                techniques: photometric --
                techniques: radial velocities --
                techniques: spectroscopic
               }

   \maketitle


\section{Introduction}

Since the discovery of the first extrasolar planet around a Solar-type star by \cite{mayor}, more than 1000 planets have been detected outside our Solar system\footnote{\url{http://exoplanet.eu/}}. Among this large harvest, the sub-sample of planets that transit the disc of their host star is extremely valuable. Indeed, transiting exoplanets allow parameters such as mass, radius and density to be accurately determined (e.g. \citealt{charbonneau}), as well as their atmospheric properties to be studied during their transits and occultations (e.g. \citealt{seager2}). At the time of writing, over 400 transiting planets have been discovered$ ^{1}$, a significant fraction of them being Jovian-type planets orbiting within 0.1 AU of their host star. Most of these so-called ``hot Jupiters'' were detected by ground-based transit surveys, among which the WASP survey (Wide Angle Search for Planets, \citealt{pollacco}) has been the most successful, with now more than 100 planets discovered (\citealt{hellier13}). Ongoing WASP discoveries are important for the field of exoplanetology as these systems tend to be particularly prone to thorough characterizations, owing to their bright host stars ($9<V<13$), short orbits and favorable planet-to-star area ratios. Therefore, they will be prime targets for thorough characterizations with future facilities such as CHEOPS (CHaracterising ExOPlanets Satellite, \citealt{cheops}) and JWST (James Webb Space Telescope, \citealt{jwst}).\\
\indent
In this paper, we report the discovery of three additional transiting planets by the WASP survey. WASP-68 b is a \hbox{0.95 $M_{\mathrm{Jup}}$} planet in a 5 days orbit around a G0-type star, WASP-73 b is a dense 1.88 $M_{\mathrm{Jup}}$ planet orbiting an F9-type star every \hbox{4.1 days}, while \hbox{WASP-88 b} is a super-bloated 0.56 $M_{\mathrm{Jup}}$ planet in a \hbox{4.9 days} orbit around an F6-type star. All three host stars appear to be significantly evolved. Consequently, they have relatively large radii ($R_{\star}$ = 1.7-2.1 $R_{\odot}$) translating into long \hbox{(5-6 hours)} and low-amplitude transits for the three planets: $\sim$0.6\% for \hbox{WASP-68 b}, $\sim$0.3\% for WASP-73 b (the shallowest transits yet for a WASP planet) and $\sim$0.7\% for WASP-88 b. Their detection demonstrates therefore the excellent photometric potential of the WASP survey.\\
\indent
Section \ref{obs} presents the WASP discovery photometry as well as the follow-up photometric and spectroscopic observations that we used to confirm and characterize the three systems. In Section \ref{analysis}, we describe the spectroscopic determination of the stellar atmospheric properties and the derivation of the systems' parameters through combined analyses of our photometric and spectroscopic data. Finally, we discuss and summarize our results in Section \ref{discussion}.

\section{Observations}
\label{obs}

\subsection{WASP transit detection photometry}

\begin{figure}[h!]
\caption{WASP photometry for WASP-68 (\textit{top}), WASP-73 (\textit{middle}) and WASP-88 (\textit{bottom}) folded on the best-fitting transit ephemeris from the transit search algorithm presented in \cite{collier}, and binned per 0.01d intervals.} 
\vspace{-2.0cm}
\centering   
\includegraphics[bb=40 427 563 687, width=0.44\textwidth]{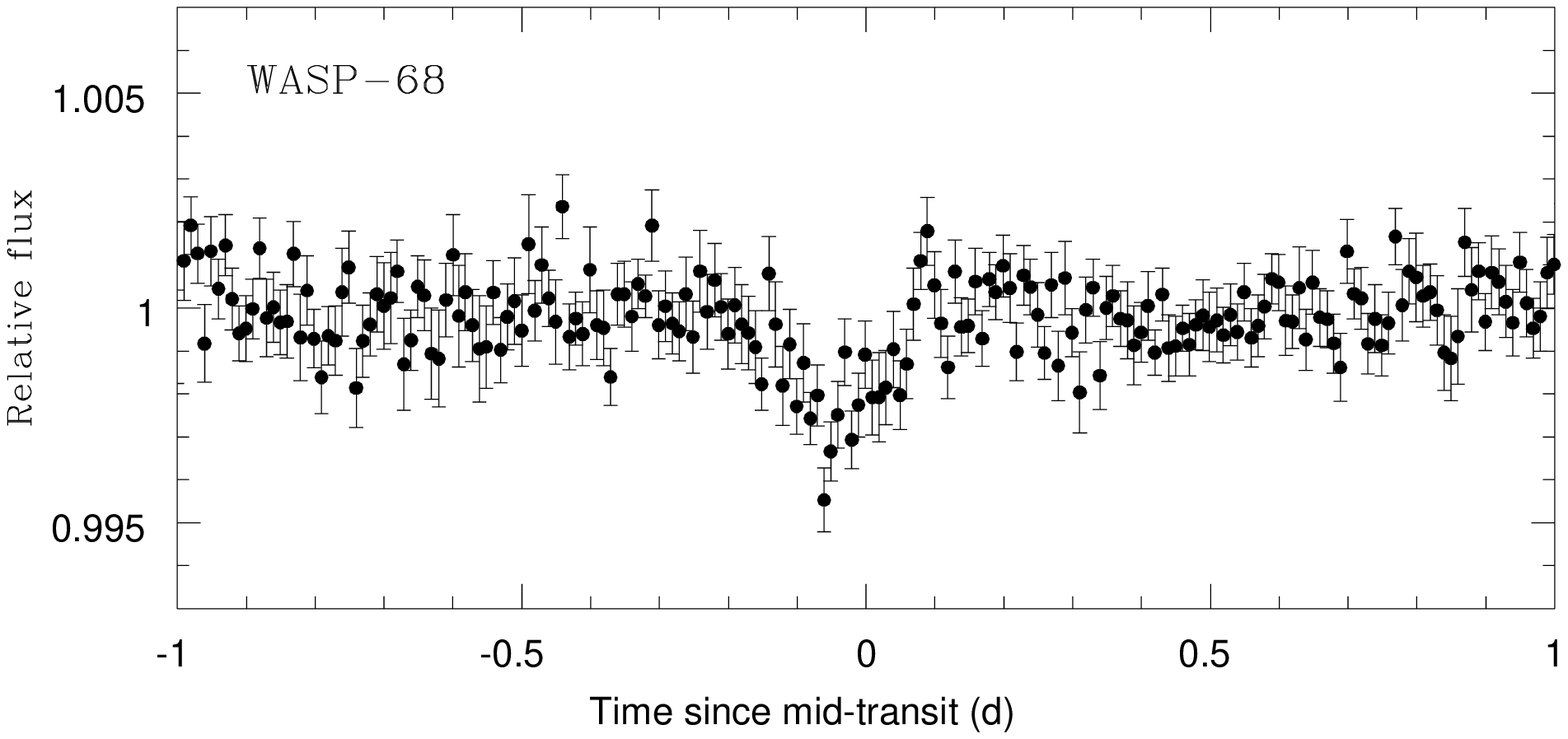}
\includegraphics[bb=40 427 563 687, width=0.44\textwidth]{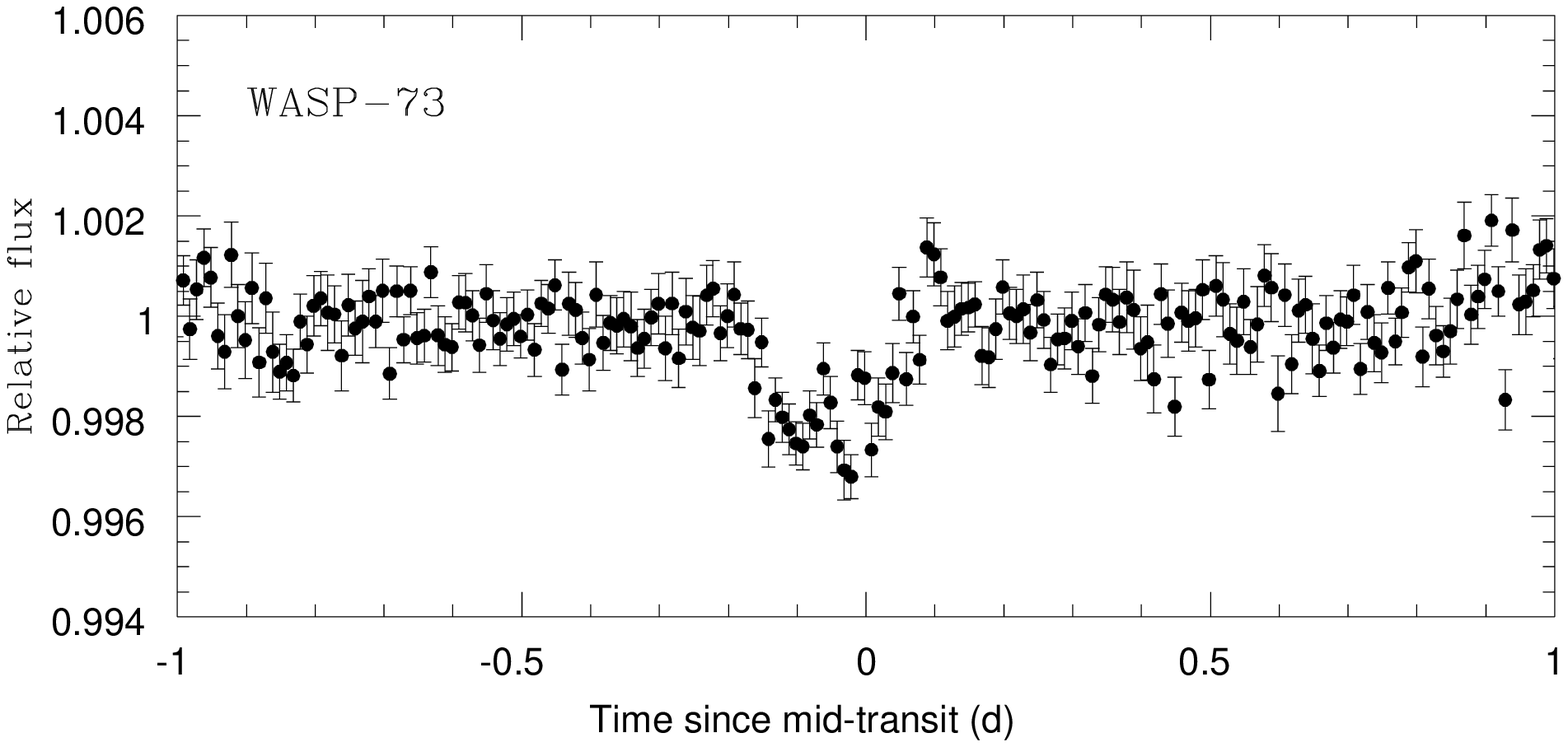}
\includegraphics[bb=40 427 563 687, width=0.44\textwidth]{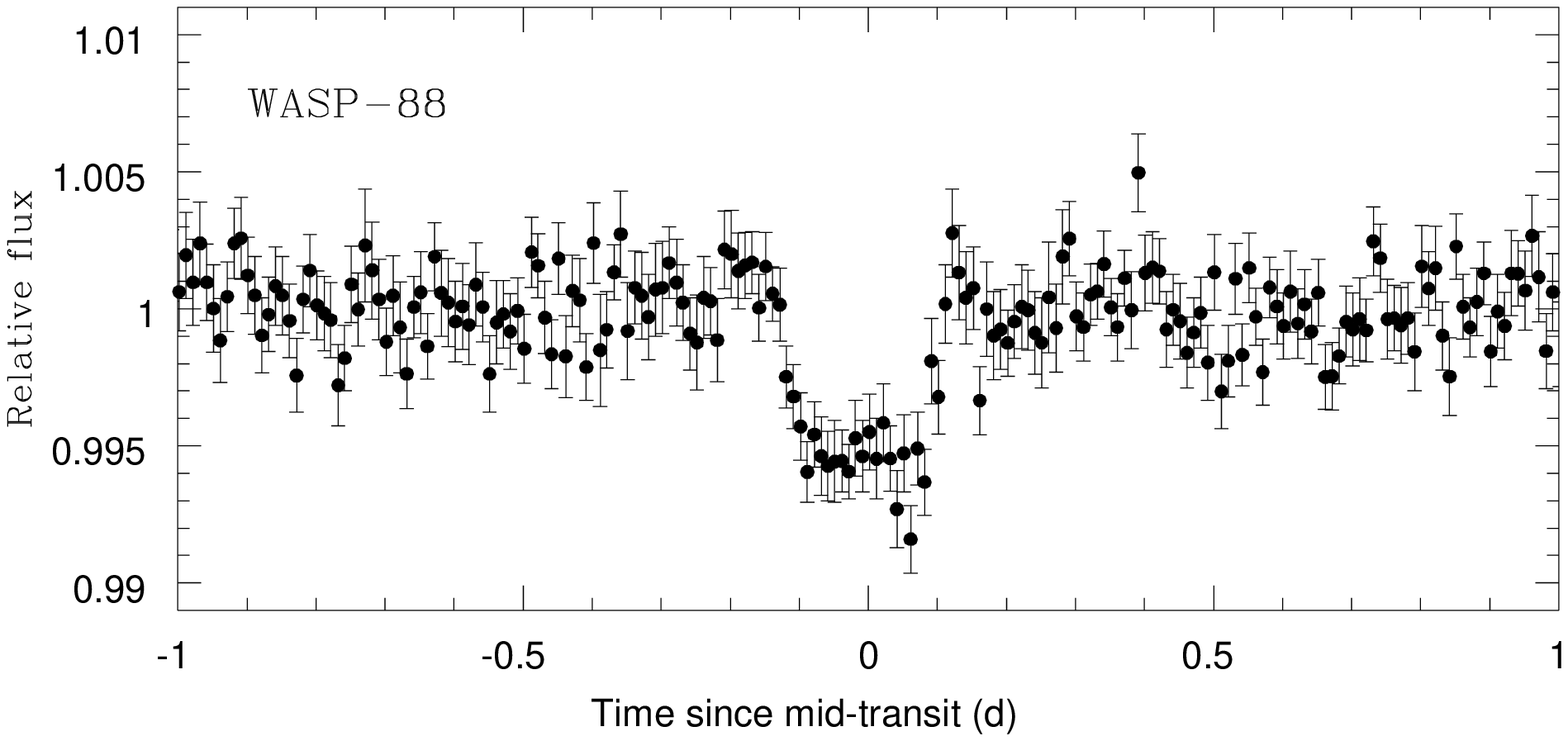}
\vspace{+1.6cm}
\label{wasp_lc}
\end{figure}

The WASP transit survey is operated from two sites, one for each hemisphere: the Observatorio del Roque de los Muchachos in the Canary Islands in the North, and the Sutherland Station of the South African Astronomical Observatory (SAAO) in the South. Each facility consists of eight Canon 200mm f/1.8 focal lenses coupled to e2v 2048$\times$2048 pixels CCDs, which yield a field of view of 450 $\mathrm{deg}^{2}$ for each site, with a corresponding pixel scale of 13.7''/pixel. Further details of the instruments, survey and data reduction procedures can be found in \cite{pollacco} while details of the candidate selection process can be found in \cite{collier} and \cite{collier2}. The three targets presented here, WASP-68 (1SWASPJ202022.98-191852.9 = 2MASS20202298-1918528, $V$=10.7, $K$=8.9), WASP-73 (1SWASPJ211947.91-580856.0 = 2MASS21194790-5808559, $V$=10.5, $K$=9.0) and WASP-88 (1SWASPJ203802.70-482743.2 = 2MASS20380268-4827434, $V$=11.4, $K$=10.3), were observed exclusively from the southern WASP site. In total, 20804 data points were obtained for WASP-68 between May 2006 and October 2011, 50588 measurements were gathered for WASP-73 between June 2008 and November 2011, while 39906 data points were obtained for WASP-88 between June 2008 and October 2011. For each target, the WASP data were processed and searched for transit signals as described in \cite{collier}, leading to the detection of periodic dimmings in the light curves of \hbox{WASP-68}, -73 and -88 with periods of 5.084 d, 4.087 d and 4.954 d, respectively. Figure \ref{wasp_lc} presents for the three objects the WASP photometry folded on the deduced transit ephemeris.\\ 
\indent
The method described in \cite{maxted} was used to search for rotational modulation in the photometry of each object, but no periodic signal was found above the mmag amplitude.

\subsection{Spectroscopy and radial velocities}

\begin{table}[t!]
\caption{CORALIE radial-velocity measurements for WASP-68 (BS = bisector spans).}
\centering
\begin{tabular}{ccccc}
  \hline
  \hline
  Target & $\mathrm{HJD_{TDB}}$- 2 450 000 & $RV$ & $\sigma_{RV}$ & BS \\
   & & (km $\mathrm{s}^{-1}$) & (m $\mathrm{s}^{-1}$) & (km $\mathrm{s}^{-1}$) \\
  \hline
  \hline
WASP-68 & 5706.775395 & 28.35403 & 7.36 & 0.02521\\	
WASP-68 & 5707.797750 & 28.39089 & 5.25 & -0.01458\\	
WASP-68 & 5713.793654 & 28.31392 & 5.99 & -0.00789\\	
WASP-68 & 5715.766225 & 28.21345 & 5.02 & 0.00389\\	
WASP-68 & 5722.816076 & 28.38649 & 4.48 & -0.01517\\	
WASP-68 & 5765.753050 & 28.22263 & 5.84 & 0.02314\\	
WASP-68 & 5767.749560 & 28.33343 & 6.34 & 0.00869\\	
WASP-68 & 5768.728542 & 28.38024 & 5.31 & -0.00094\\	
WASP-68 & 5769.731770 & 28.32243 & 4.62 & 0.00260\\	
WASP-68 & 5770.769094 & 28.21813 & 5.56 & 0.01505\\	
WASP-68 & 5772.795194 & 28.33309 & 14.45 & -0.00200\\	
WASP-68 & 5777.773957 & 28.32040 & 5.29 & 0.01429\\	
WASP-68 & 5794.600961 & 28.38663 & 5.17 & 0.01427\\	
WASP-68 & 5796.549753 & 28.20522 & 5.60 & 0.00768\\	
WASP-68 & 5806.645082 & 28.20412 & 5.91 & 0.03023\\	
WASP-68 & 5807.705562 & 28.25392 & 6.29 & 0.00704\\	
WASP-68 & 5823.546584 & 28.32236 & 6.42 & -0.00127\\	
WASP-68 & 5824.573366 & 28.40896 & 5.36 & 0.02367\\	
WASP-68 & 5826.655256 & 28.23628 & 7.43 & 0.01630\\	
WASP-68 & 5852.532499 & 28.19375 & 6.28 & -0.01225\\	
WASP-68 & 5856.597408 & 28.28718 & 5.57 & -0.00265\\	
WASP-68 & 5858.529438 & 28.25969 & 7.73 & -0.01544\\
WASP-68 & 5864.564865 & 28.35191 & 5.26 & -0.00650\\ 
WASP-68 & 5883.548579 & 28.23114 & 6.37 & 0.02525\\	
WASP-68 & 6021.899684 & 28.32750 & 5.37 & -0.01018\\	
WASP-68 & 6022.905102 & 28.41331 & 5.62 & -0.01400\\	
WASP-68 & 6048.914901 & 28.37000 & 5.65 & -0.00286\\	
WASP-68 & 6067.850937 & 28.35299 & 4.79 & -0.00075\\	
WASP-68 & 6076.878089 & 28.21947 & 8.57 & -0.02511\\	
WASP-68 & 6103.663851 & 28.37654 & 5.03 & 0.00509\\	
WASP-68 & 6130.635583 & 28.36594 & 8.71 & 0.04089\\	
WASP-68 & 6135.775843 & 28.35722 & 6.25 & -0.01097\\	
WASP-68 & 6150.631492 & 28.38702 & 6.00 & 0.00044\\	
WASP-68 & 6183.679884 & 28.22057 & 4.90 & 0.01819\\	
WASP-68 & 6184.656200 & 28.34522 & 5.05 & -0.00802\\	
WASP-68 & 6204.554305 & 28.29089 & 7.49 & -0.00338\\	
WASP-68 & 6216.592455 & 28.40205 & 6.16 & 0.00275\\	
WASP-68 & 6431.931490 & 28.23109 & 6.12 & -0.00857\\	
WASP-68 & 6451.773166 & 28.26579 & 4.88 & -0.02323\\	
WASP-68 & 6475.858923 & 28.41736 & 6.31 & 0.02107\\	
WASP-68 & 6485.724714 & 28.43317 & 5.45 & -0.00978\\	
WASP-68 & 6507.676974 & 28.27546 & 6.20 & 0.01077\\	
WASP-68 & 6530.605288 & 28.38629 & 7.38 & 0.02674\\	
\hline
\hline
\end{tabular}
\vspace{-0.7cm}
\label{rvs68}
\end{table}

\begin{table}[t!]
\caption{CORALIE radial-velocity measurements for WASP-73 (BS = bisector spans).}
\centering
\begin{tabular}{ccccc}
  \hline
  \hline
  Target & $\mathrm{HJD_{TDB}}$- 2 450 000 & $RV$ & $\sigma_{RV}$ & BS \\
   & & (km $\mathrm{s}^{-1}$) & (m $\mathrm{s}^{-1}$) & (km $\mathrm{s}^{-1}$) \\
  \hline
  \hline
WASP-73 & 5835.606190 & 10.44179 & 9.03 & 0.03075\\	  
WASP-73 & 5837.502076 & 10.82294 & 7.06 & 0.03546\\	  	  
WASP-73 & 5851.645089 & 10.41902 & 7.43 & -0.02858\\	  
WASP-73 & 5856.643946 & 10.54277 & 7.65 & 0.05603\\	  	  
WASP-73 & 5858.553071 & 10.75127 & 9.64 & 0.02465\\	  
WASP-73 & 5859.533285 & 10.45047 & 7.41 & 0.06098\\
WASP-73 & 5864.601220 & 10.48392 & 7.12 & 0.03086\\
WASP-73 & 5865.620159 & 10.76933 & 8.54 & 0.05479\\
WASP-73 & 5880.549374 & 10.44138 & 7.82 & -0.00327\\	  
WASP-73 & 5892.578470 & 10.45665 & 12.22 & 0.04721\\	  
WASP-73 & 5893.565583 & 10.60197 & 7.77 & -0.00459\\	  
WASP-73 & 5894.530903 & 10.80197 & 7.37 & 0.04810\\	  
WASP-73 & 6130.697716 & 10.60558 & 14.86 & 0.01835\\	 
WASP-73 & 6137.897482 & 10.43505 & 10.11 & 0.00469\\	  
WASP-73 & 6149.772409 & 10.48072 & 8.69 & 0.03588\\	  
WASP-73 & 6158.765097 & 10.48112 & 8.69 & 0.06660\\	  
WASP-73 & 6216.615675 & 10.64684 & 8.74 & 0.04270\\	  
WASP-73 & 6488.833899 & 10.50459 & 8.52 & -0.02173\\	  
WASP-73 & 6546.740705 & 10.41467 & 8.73 & 0.05445\\	  
WASP-73 & 6547.703915 & 10.64815 & 7.11 & 0.02528\\	  
\hline
\hline
\end{tabular}
\vspace{-0.5cm}
\label{rvs73}
\end{table} 


\begin{table}[t!]
\centering
\vspace{0.2cm}
\caption{CORALIE radial-velocity measurements for WASP-88 (BS = bisector spans).}
\begin{tabular}{ccccc}
  \hline
  \hline
  Target & $\mathrm{HJD_{TDB}}$- 2 450 000 & $RV$ & $\sigma_{RV}$ & BS \\
   & & (km $\mathrm{s}^{-1}$) & (m $\mathrm{s}^{-1}$) & (km $\mathrm{s}^{-1}$) \\
  \hline
  \hline
WASP-88 & 5834.557656 & -6.75688 & 17.97 & 0.11267\\   
WASP-88 & 5856.620397 & -6.82192 & 24.19 & 0.13488\\   
WASP-88 & 6119.775893 & -6.80810 & 17.12 & 0.00269\\   
WASP-88 & 6121.751740 & -6.72092 & 19.50 & 0.11340\\   
WASP-88 & 6123.632372 & -6.81181 & 19.96 & 0.02430\\   
WASP-88 & 6124.600910 & -6.86256 & 17.64 & 0.03914\\   
WASP-88 & 6125.676195 & -6.77344 & 17.17 & -0.03997\\  	
WASP-88 & 6133.736344 & -6.80539 & 20.57 & 0.00683\\	  
WASP-88 & 6134.833084 & -6.79385 & 30.92 & 0.03759\\   
WASP-88 & 6135.804118 & -6.76124 & 22.68 & -0.02308\\  
WASP-88 & 6136.820069 & -6.74445 & 20.80 & -0.02308\\  
WASP-88 & 6137.869386 & -6.77739 & 23.96 & 0.11208\\   
WASP-88 & 6149.574773 & -6.80985 & 22.88 & 0.03215\\   
WASP-88 & 6150.543194 & -6.73250 & 25.12 & 0.05769\\   
WASP-88 & 6154.586060 & -6.76785 & 24.99 & 0.06009\\   
WASP-88 & 6172.627045 & -6.84122 & 17.86 & -0.00501\\  
WASP-88 & 6173.706647 & -6.87322 & 36.70 & 0.02791\\   
WASP-88 & 6475.892228 & -6.83574 & 22.68 & 0.02641\\   
WASP-88 & 6480.804868 & -6.82613 & 29.60 & 0.05213\\   
WASP-88 & 6487.935947 & -6.70113 & 21.36 & 0.08684\\   
WASP-88 & 6558.499522 & -6.76226 & 20.15 & 0.05448\\
WASP-88 & 6563.582029 & -6.71375 & 30.03 & -0.02524\\
WASP-88 & 6567.558431 & -6.70463 & 17.69 & 0.03362\\
\hline
\hline
\end{tabular}
\label{rvs88}
\vspace{-0.5cm}
\end{table}

\begin{figure}[t!]
\centering   
\caption{\textit{Top:} CORALIE RVs for WASP-68 phase-folded on the best-fit orbital period, with the best-fit Keplerian model over-imposed in red. \textit{Bottom:} correlation diagram CCF bisector span vs. RV. The colors indicate the measurement timings.} 
\vspace{0.3cm}
\includegraphics[bb=0 420 563 550, angle=0, width=0.45\textwidth]{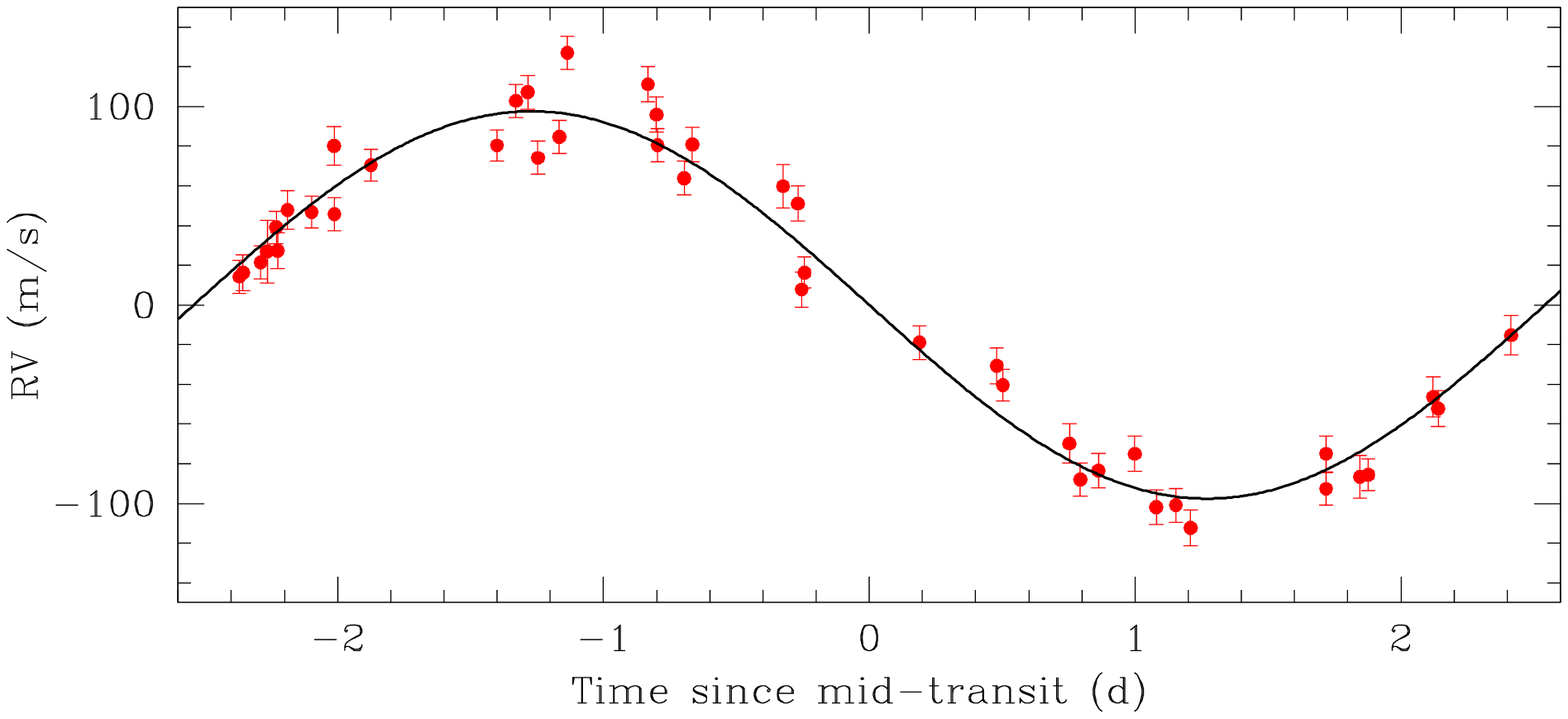} 
\includegraphics[width=0.53\textwidth]{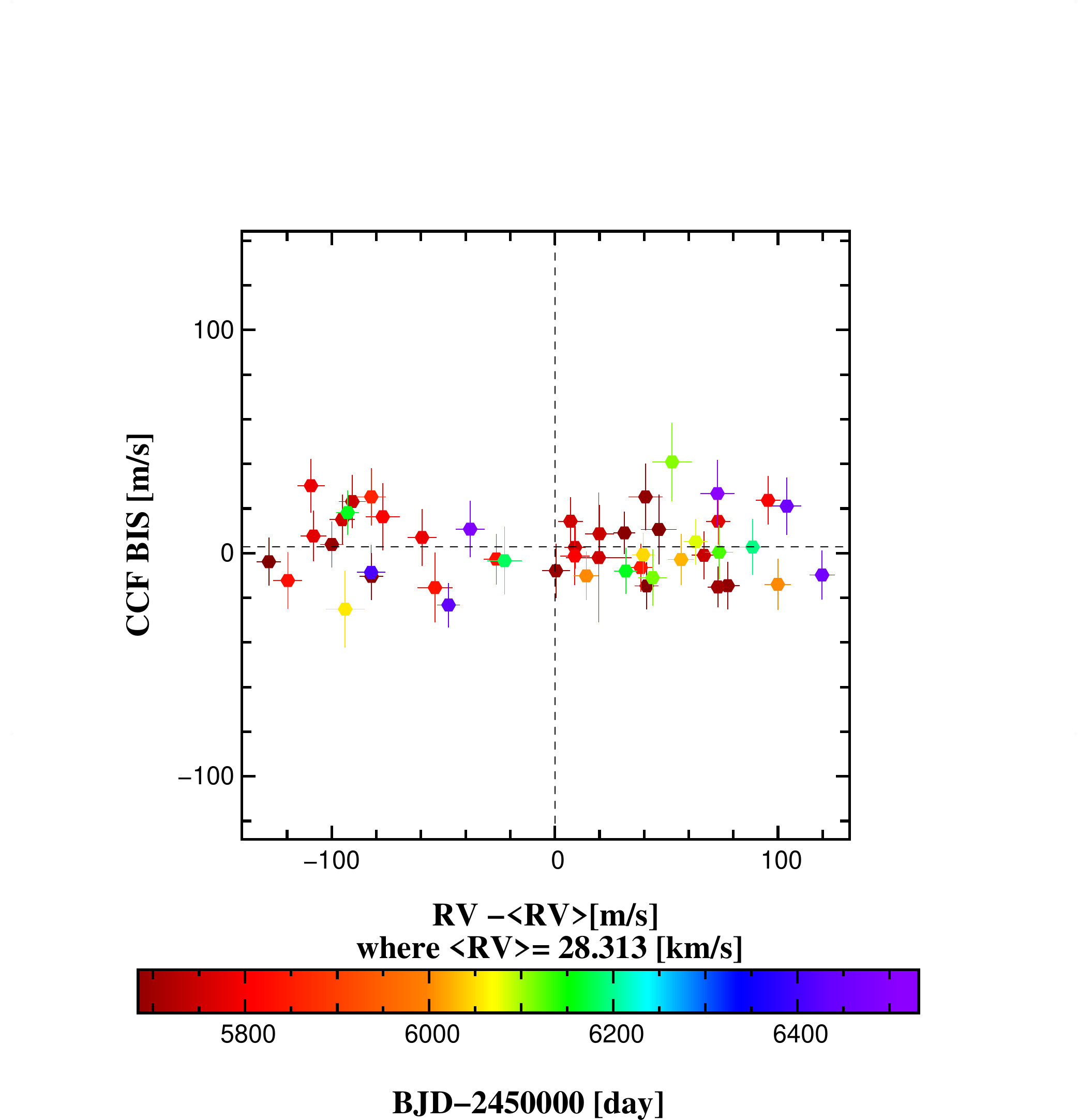}              
\label{plotrv68}
\end{figure}

\begin{figure}[h!]
\caption{\textit{Top:} CORALIE RVs for WASP-73 phase-folded on the best-fit orbital period, with the best-fit Keplerian model over-imposed in red. \textit{Bottom:} correlation diagram CCF bisector span vs. RV. The colors indicate the measurement timings.} 
\vspace{0.3cm}
\centering   
\includegraphics[bb=0 420 563 550, angle=0, width=0.45\textwidth]{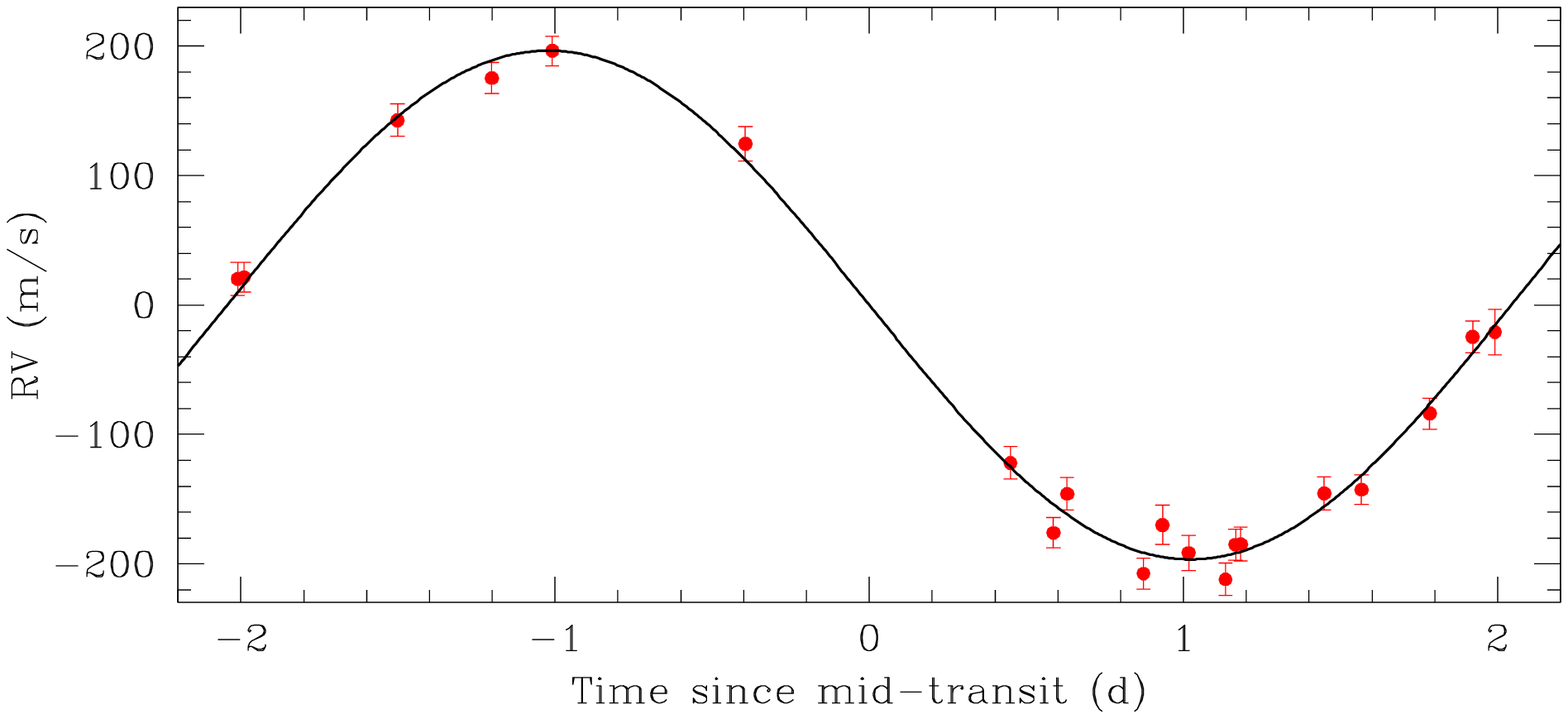} 
\includegraphics[width=0.53\textwidth]{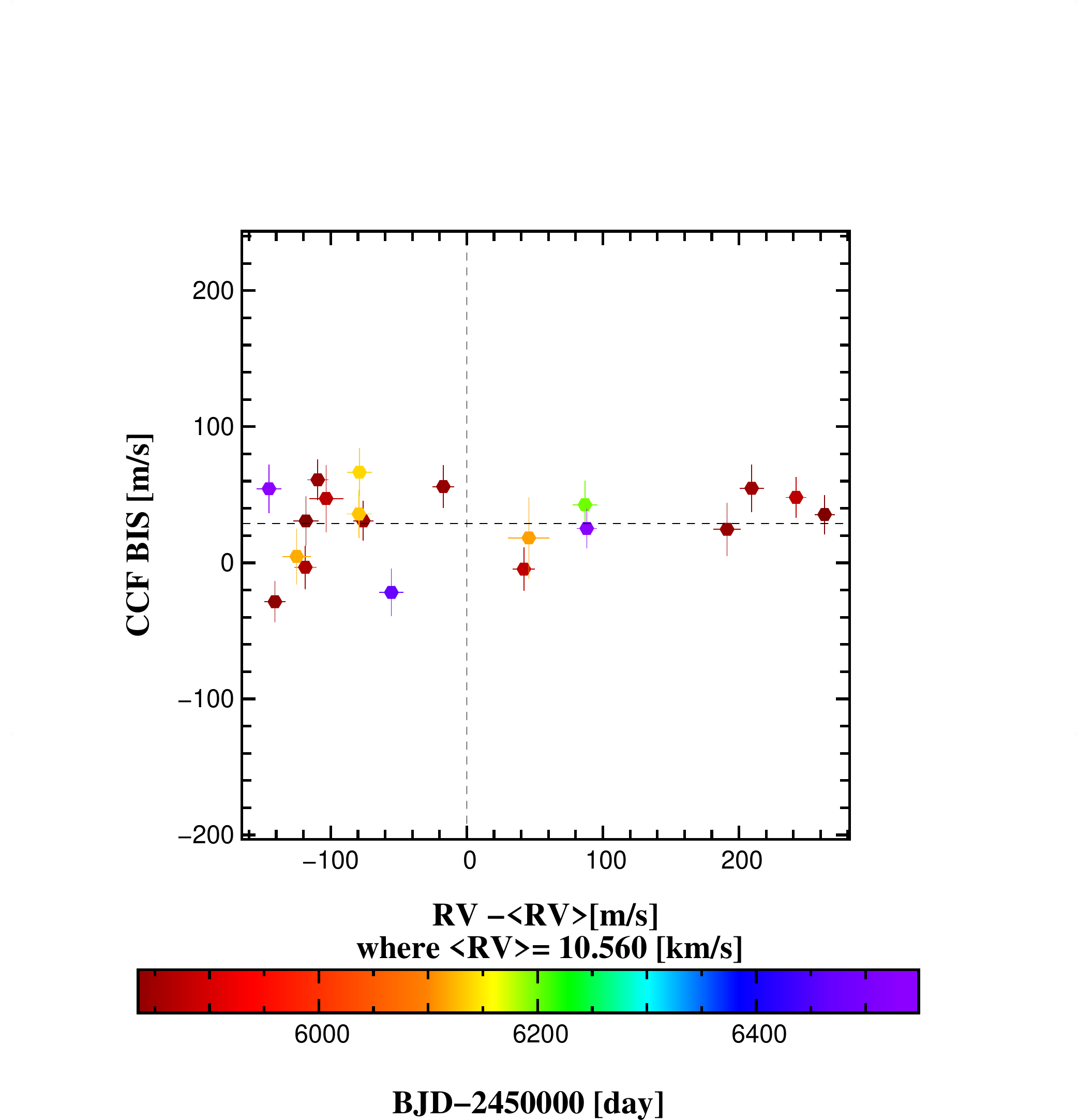}              
\label{plotrv73}
\end{figure}

\begin{figure}[h!]
\caption{\textit{Top:} CORALIE RVs for WASP-88 phase-folded on the best-fit orbital period, with the best-fit Keplerian model over-imposed in red. \textit{Bottom:} correlation diagram CCF bisector span vs. RV. The colors indicate the measurement timings.} 
\centering   
\vspace{0.3cm}
\includegraphics[bb=0 420 563 550, angle=0, width=0.45\textwidth]{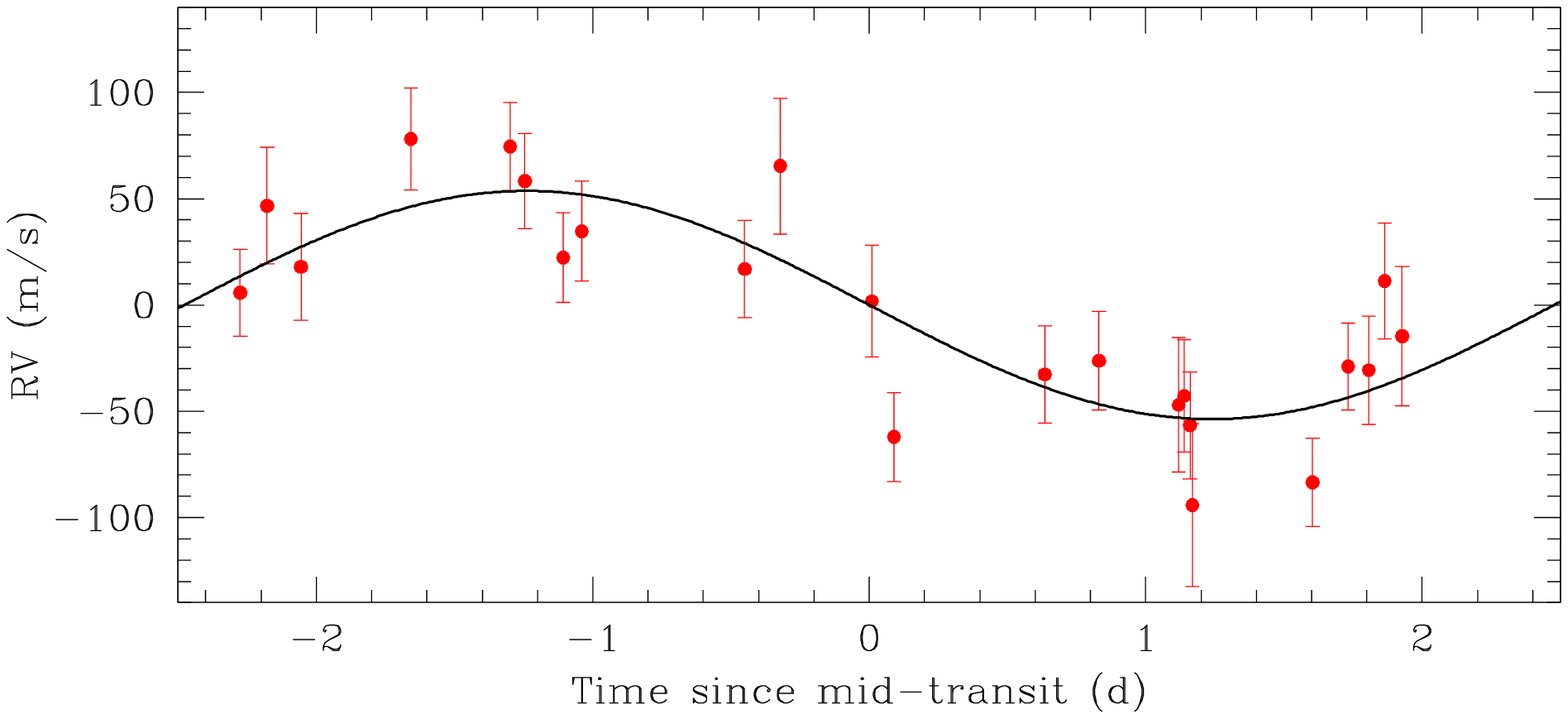} 
\includegraphics[width=0.53\textwidth]{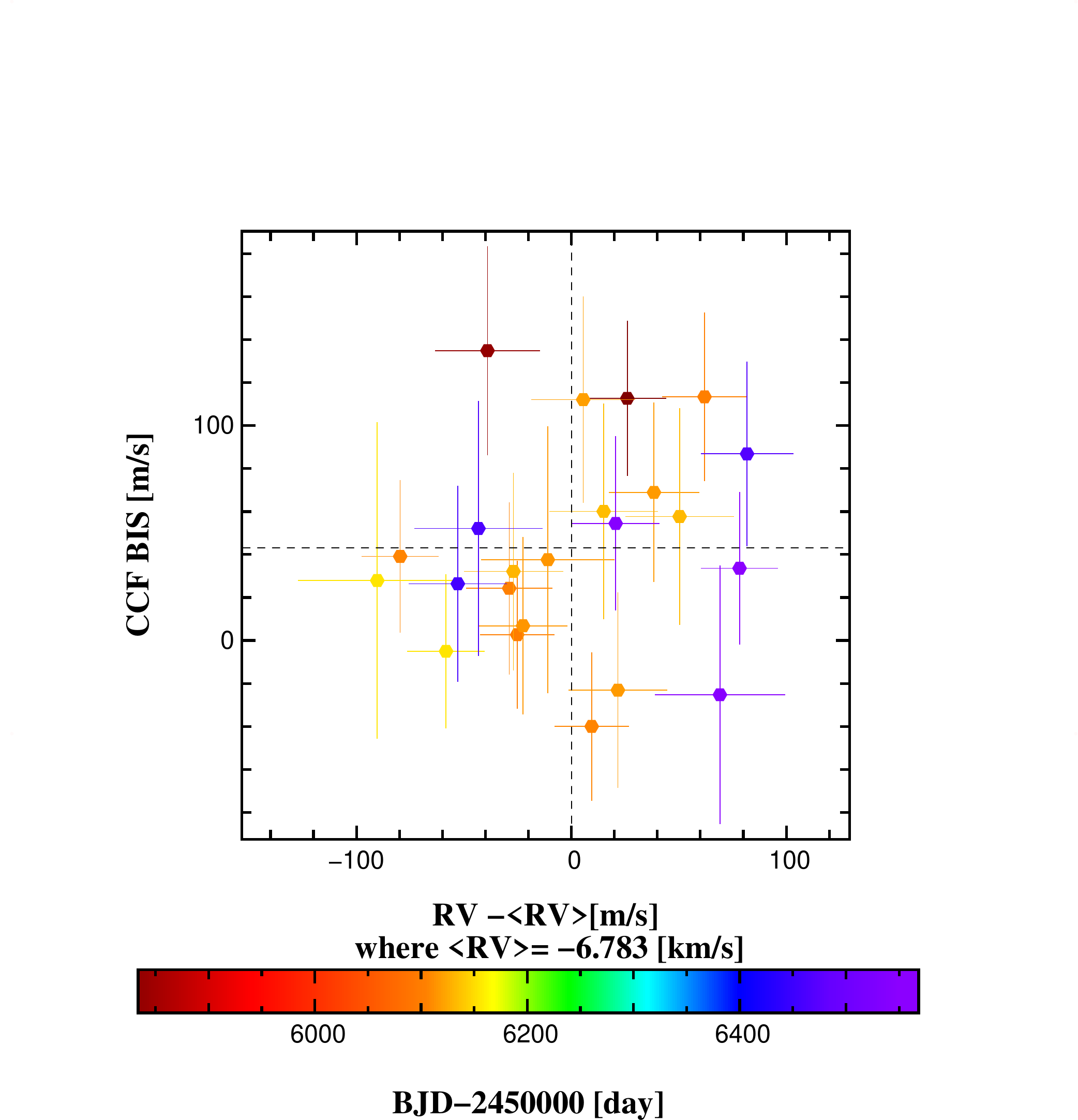}              
\label{plotrv88}
\end{figure}

Spectroscopic measurements of each star were obtained using the CORALIE spectrograph mounted on the 1.2m Euler-Swiss telescope at the La Silla site (Chile). 43 spectra were gathered for WASP-68 between May 2011 and August 2013, 20 spectra were obtained for WASP-73 from October 2011 to September 2013, while 23 spectra were gathered for WASP-88 between September 2011 and October 2013. For all spectroscopic observations, radial velocities (RVs) were computed using the weighted cross-correlation technique described in \cite{baranne}. These RVs are presented in Tables \ref{rvs68}, \ref{rvs73} and \ref{rvs88}. For each star, RV variations were detected with periods similar to those found in the WASP photometry and with semi-amplitudes consistent with planetary-mass companions (see Figures \ref{plotrv68}, \ref{plotrv73} and \ref{plotrv88}, upper panels).\\ 
\indent
In order to discard any false-positive scenarios that could create RV variations mimicking planetary signatures, we checked the CORALIE cross-correlation functions (CCF) bisector spans according to the technique described by \cite{queloz}. Indeed, false positives such as blended eclipsing binaries or starspots would also induce spectral-line distortions, resulting in correlated variations of RVs and bisector spans. This effect was for example observed for the HD41004 system (\citealt{santos}), which consists of a K-dwarf blended with an M-dwarf companion (separation $\sim$0.5'') orbited itself by a short-period brown dwarf. For this extreme system, the RVs showed a clear signal at the period of the brown dwarf orbit (1.3 d) and with an amplitude $\sim$50 m $\mathrm{s}^{-1}$ that could have been interpreted as the signal of a sub-Saturn mass planet orbiting the K-dwarf. However, the 0.67$\pm$0.03 slope of the RV-bisector relation clearly revealed that the observed signal did not originate from the K-dwarf and shed light on the blended nature of the system.\\
\indent 
For our three systems, the bisector spans revealed to be stable, their standard deviation being close to their average error (15 vs 12 m $\mathrm{s}^{-1}$ for WASP-68, 27 vs 18 m $\mathrm{s}^{-1}$ for WASP-73 and 48 vs 45 m $\mathrm{s}^{-1}$ for WASP-88). No correlation between the RVs and the bisector spans was found with $p$-value less than 0.05 (see Figures \ref{plotrv68}, \ref{plotrv73} and \ref{plotrv88}, lower panels), the slopes deduced from linear regressions being -0.01$\pm$0.03 (WASP-68), 0.04$\pm$0.05 (WASP-73) and 0.10$\pm$0.21 (WASP-88). These values and errors support our conclusion that the periodic dimming and RV variation of each system are well-caused by a transiting planet. This conclusion is also strengthened by the consistency of the solutions derived from the global analysis of our spectroscopic and photometric data (see Section \ref{mcmc}).

\vspace{-0.2cm}
\subsection{Follow-up photometry}

In order to refine the systems' parameters, high-quality transit observations were obtained using the 0.6m TRAPPIST robotic telescope (TRAnsiting Planets and PlanetesImals Small Telescope) and the EulerCam CCD camera mounted on the 1.2m Euler-Swiss telescope, both located at ESO La Silla Observatory. These follow-up light curves are summarized in Table \ref{obstable} and presented in Figures \ref{lcs68}, \ref{lcs73} and \ref{lcs88}.

\begin{table*}[t!]
\centering
\caption{Summary of follow-up photometry obtained for WASP-68, WASP-73 and WASP-88. For each light curve, this table shows the date of acquisition, the used instrument and filter, the number of data points, the selected baseline function, the standard deviation of the best-fit residuals (unbinned and binned per intervals of 2 min), and the deduced values for $\beta_{w}$, $\beta_{r}$ and $CF=\beta_{w} \times \beta_{r}$. For the baseline function, p($\epsilon^{N}$) denotes, respectively, a $N$-order polynomial function of time ($\epsilon=t$), airmass ($\epsilon=a$), PSF full-width at half maximum ($\epsilon=f$), background ($\epsilon=b$), and $x$ and $y$ positions ($\epsilon=xy$). $o$ denotes an offset fixed at the time of the meridian flip.}
\vspace{0.2cm}
\begin{tabular}{cccccccccccc}
  \hline
  \hline 
 Target & Night & Telescope & Filter & $N_{p}$ & $T_{\mathrm{exp}}$ & Baseline function & $\sigma$ & $\sigma_{120\mathrm{s}}$ & $\beta_{w}$ & $\beta_{r}$ & $CF$  \\
  & & & & & (s) & & (\%) & (\%) & & & \\
  \hline
  WASP-68 & 2012 May 16-17 & TRAPPIST & $I+z$ & 1139 & 8 & p($a^{1}$+$f^{1}$+$xy^{1}$) + $o$ & 0.28 & 0.10 & 1.03 & 1.22 & 1.26\\
  WASP-68 & 2012 Jul. 06-07 & TRAPPIST & $I+z$ & 1181 & 8 & p($a^{2}$+$b^{1}$+$xy^{2}$) + $o$ & 0.30 & 0.11 & 1.21 & 1.27 & 1.54\\
  WASP-68 & 2013 Jul. 02-03 & TRAPPIST & $I+z$ & 1357 & 8 & p($t^{2}$+$b^{1}$+$xy^{1}$) + $o$ & 0.25 & 0.10 & 0.88 & 1.57 & 1.39\\
  WASP-68 & 2012 Jul. 06-07 & EulerCam & $I_{c}$ & 412 & 40 & p($t^{2}$+$f^{2}$+$b^{1}$) & 0.10 & 0.07 & 1.40 & 1.00 & 1.40\\
  WASP-68 & 2013 Jul. 02-03 & EulerCam & $I_{c}$ & 333 & 50 & p($a^{1}$+$f^{1}$+$xy^{1}$) & 0.10 & 0.08 & 1.36 & 2.15 & 2.93\\ 
   & & & & & & & & & & & \\
  WASP-73 & 2012 Jul. 19-20 & TRAPPIST & $z'$ & 476 & 25 & p($t^{2}$) & 0.22 & 0.12 & 1.13 & 1.08 & 1.22\\
  WASP-73 & 2012 Jul. 23-24 & TRAPPIST & $z'$ & 640 & 25 & p($a^{1}$+$f^{1}$) + $o$ & 0.25 & 0.14 & 1.47 & 1.11 & 1.63\\
  WASP-73 & 2012 Jul. 19-20 & EulerCam & Gunn-$r'$ & 346 & 60 & p($t^{2}$+$f^{2}$+$xy^{1}$) & 0.10 & 0.07 & 1.48 & 1.54 & 2.27\\
  WASP-73 & 2012 Jul. 23-24 & EulerCam & Gunn-$r'$ & 352 & 50 & p($a^{1}$+$xy^{1}$) & 0.13 & 0.10 & 1.61 & 2.55 & 4.11\\  
   & & & & & & & & & & & \\
  WASP-88 & 2012 Aug. 27-28 & TRAPPIST & $I+z$ & 867 & 20 & p($a^{1}$) + $o$ & 0.23 & 0.11 & 1.10 & 1.19 & 1.31\\
  WASP-88 & 2013 Jun. 30-Jul. 01 & TRAPPIST & $I+z$ & 837 & 20 & p($t^{2}$) + $o$ & 0.37 & 0.21 & 1.59 & 1.14 & 1.82\\
  WASP-88 & 2012 Aug. 12-13 & EulerCam & Gunn-$r'$ & 95 & 80 & p($a^{1}$+$xy^{2}$) & 0.08 & 0.08 & 1.25 & 1.00 & 1.25\\
  WASP-88 & 2013 Jun. 25-26 & EulerCam & Gunn-$r'$ & 246 & 70 & p($a^{1}$+$f^{1}$) & 0.09 & 0.09 & 1.31 & 2.33 & 3.05\\
  WASP-88 & 2013 Jun. 30-Jul. 01 & EulerCam & Gunn-$r'$ & 317 & 70 & p($t^{2}$+$b^{1}$) & 0.17 & 0.17 & 2.45 & 1.08 & 2.65\\
  \hline
  \hline
\end{tabular}
\vspace{0.3cm}
\label{obstable}
\end{table*}

\subsubsection{TRAPPIST observations} 

TRAPPIST is a 60cm robotic telescope dedicated to the detection and characterization of transiting exoplanets and to the photometric monitoring of bright comets and other small bodies. It is equipped with a thermoelectrically-cooled 2K$\times$2K CCD having a pixel scale of 0.65'' that translates into a 22'$\times$22' field of view. For details of TRAPPIST see \cite{gillon} and \cite{jehin}. The TRAPPIST photometry was obtained using a readout mode of 2$\times$2 MHz with 1$\times$1 binning, resulting in a readout + overhead time of 6.1 s and a readout noise of 13.5 $\mathrm{e}^{-}$. A slight defocus was applied to the telescope in order to improve the duty cycle, spread the light over more pixels and thereby improve the sampling of the PSF. Three transits of WASP-68 b and two transits of WASP-88 b were observed through a special ``$I+z$'' filter that has a transmittance $>$90\% from 750 nm to beyond 1100 nm\footnote{\url{http://www.astrodon.com/products/filters/near-infrared/}}. For WASP-73b, two transits were observed in a Sloan $z'$ filter ($\lambda_{\mathrm{eff}}=915.9\pm0.5$ nm). During the runs, the positions of the stars on the chip were maintained to within a few pixels thanks to a ``software guiding'' system that regularly derives an astrometric solution for the most recently acquired image and sends pointing corrections to the mount if needed. After a standard pre-reduction (bias, dark, flatfield correction), the stellar fluxes were extracted from the images using the IRAF/DAOPHOT\footnote{IRAF is distributed by the National Optical Astronomy Observatory, which is operated by the Association of Universities for Research in Astronomy, Inc., under cooperative agreement with the National Science Foundation.} aperture photometry software (\citealt{stetson}). For each light curve, we tested several sets of reduction parameters and kept the one giving the most precise photometry for the stars of similar brightness as the target. After a careful selection of reference stars, the transit light curves were finally obtained using differential photometry.


\subsubsection{EulerCam observations}

EulerCam is an E2V 4K$\times$4K back-illuminated deep-depletion CCD detector installed at the Cassegrain focus of the 1.2m Euler-Swiss telescope. The field of view of EulerCam is 15.7'$\times$15.7', producing a pixel scale of 0.23''. In order to keep the stars on the same locations on the detector during the observations, EulerCam employs an ``Absolute Tracking" system very similar to the one of TRAPPIST, which matches the point sources in each image with a catalogue, and if needed, adjusts the telescope pointing between exposures to compensate for drifts. Two transits of WASP-73 b and three transits of WASP-88 b were observed with EulerCam through Gunn-$r'$ filter ($\lambda_{\mathrm{eff}}=620.4\pm0.5$ nm), while 2 transits of WASP-68 b were observed in an $I_{c}$ filter ($\lambda_{\mathrm{eff}}=806\pm0.5$ nm). Here too, a slight defocus was applied to the telescope to optimize the observation efficiency and to minimize pixel-to-pixel effects. The reduction procedure used to extract the transit light curves was similar to that performed on TRAPPIST data. Further details of the EulerCam instrument and data reduction procedures can be found in \cite{lendl}.

\vspace{-0.2cm}

\section{Analysis}
\label{analysis}

\subsection{Spectroscopic analysis - stellar atmospheric properties}
\label{barry}

For each star, the individual CORALIE spectra were co-added to produce a single spectrum with typical S/N of around 100:1. The stellar atmospheric parameters were then derived using the methods given in \cite{amanda}. These parameters are listed in Tables \ref{w68table}, \ref{w73table} and \ref{w88table} for WASP-68, WASP-73 and WASP-88, respectively. The excitation balance of the Fe~{\sc i} lines was used to determine the effective temperature $T_{\mathrm{eff}}$. The surface gravity log $g_{\star}$ was determined from the ionisation balance of Fe~{\sc i} and Fe~{\sc ii}. The Ca~{\sc i} line at 6439 {\AA} and the Na~{\sc i} D lines were also used as log $g_{\star}$ diagnostics. The elemental abundances were determined from equivalent width measurements of several unblended lines. Iron abundances are relative to the solar values obtained by \cite{asplund}. Values for microturbulence ($\xi_{\mathrm{t}}$) were determined from Fe~{\sc i} using the method of \cite{magain}. The quoted error estimates include that given by the uncertainties in $T_{\mathrm{eff}}$ and log $g_{\star}$, as well as the scatter due to measurement and atomic data uncertainties. The projected stellar rotation velocity $v$ sin $i_{\star}$ was determined by fitting the profiles of several unblended Fe~{\sc i} lines. An instrumental FWHM of 0.11 $\pm$ 0.01~{\AA} was determined for the three stars from the telluric lines around \hbox{6300 \AA}. Macroturbulence ($v_{\mathrm{mac}}$) values were obtained from the calibration of \cite{bruntt}. Spectral types were estimated from $T_{\mathrm{eff}}$ using the table in \cite{gray}. Finally, we also used the \cite{torres} calibration to obtain first stellar mass estimates : 1.24 $\pm$ 0.10 $M_{\odot}$ for WASP-68, 1.40 $\pm$ 0.12 $M_{\odot}$ for WASP-73 and 1.38 $\pm$ 0.12 $M_{\odot}$ for WASP-88.

\begin{figure}[h!]
\centering   
\caption{\textit{Top:} Individual follow-up transit light curves for \hbox{WASP-68 b}. \textit{Bottom:} Combined follow-up photometry for \hbox{WASP-68 b}. The observations are shown as red points (bin width=2min) and are period-folded on the best-fit transit ephemeris. Each light curve has been divided by the respective photometric baseline model (see Section \ref{mcmc}). For each filter, the superimposed, solid, black line is our best-fit transit model. The light curves are shifted along the \textit{y}-axis for clarity.}
\includegraphics[angle=0, width=0.48\textwidth, height=10cm]{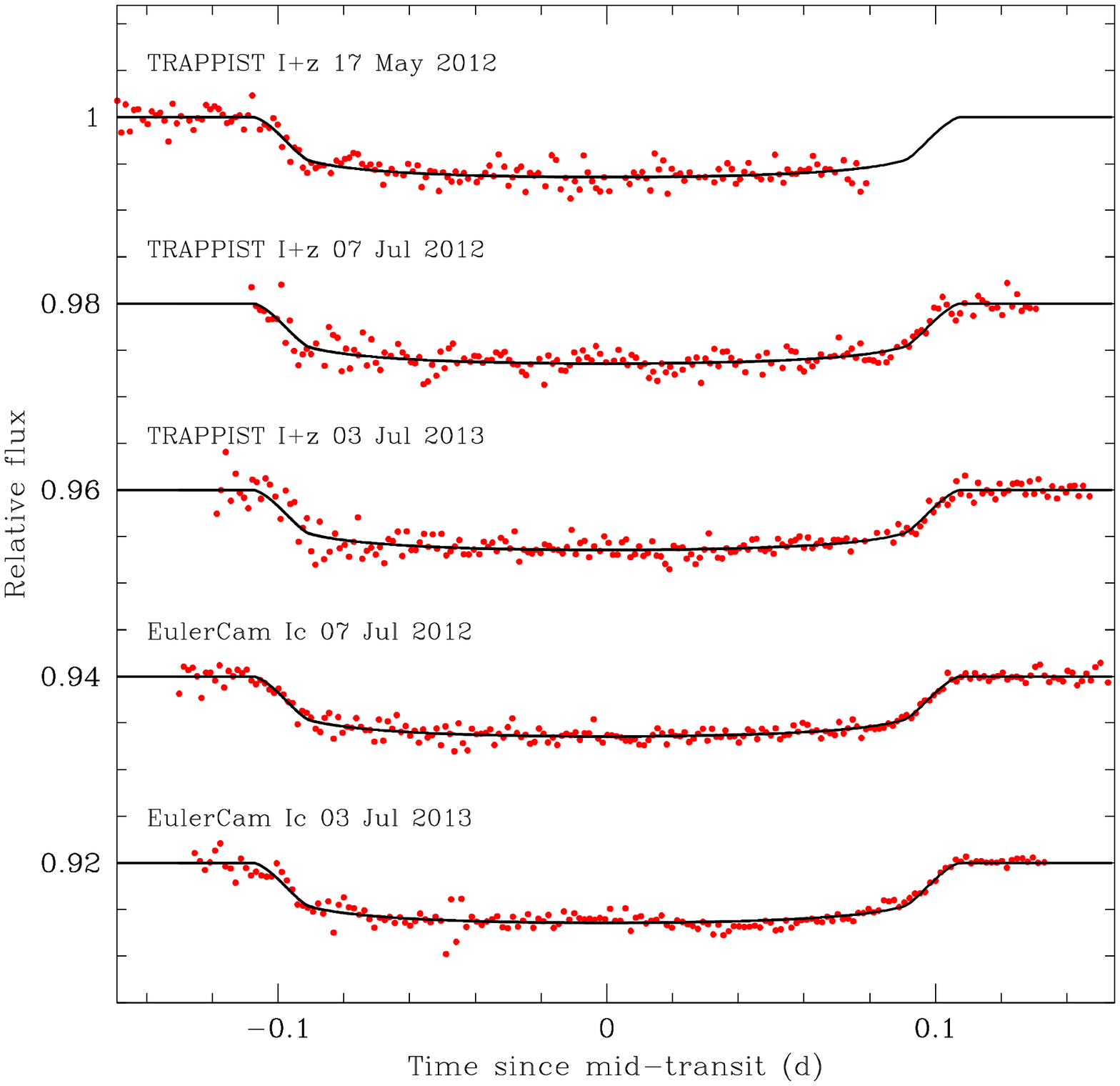}  
\includegraphics[angle=0, width=0.48\textwidth, height=10cm]{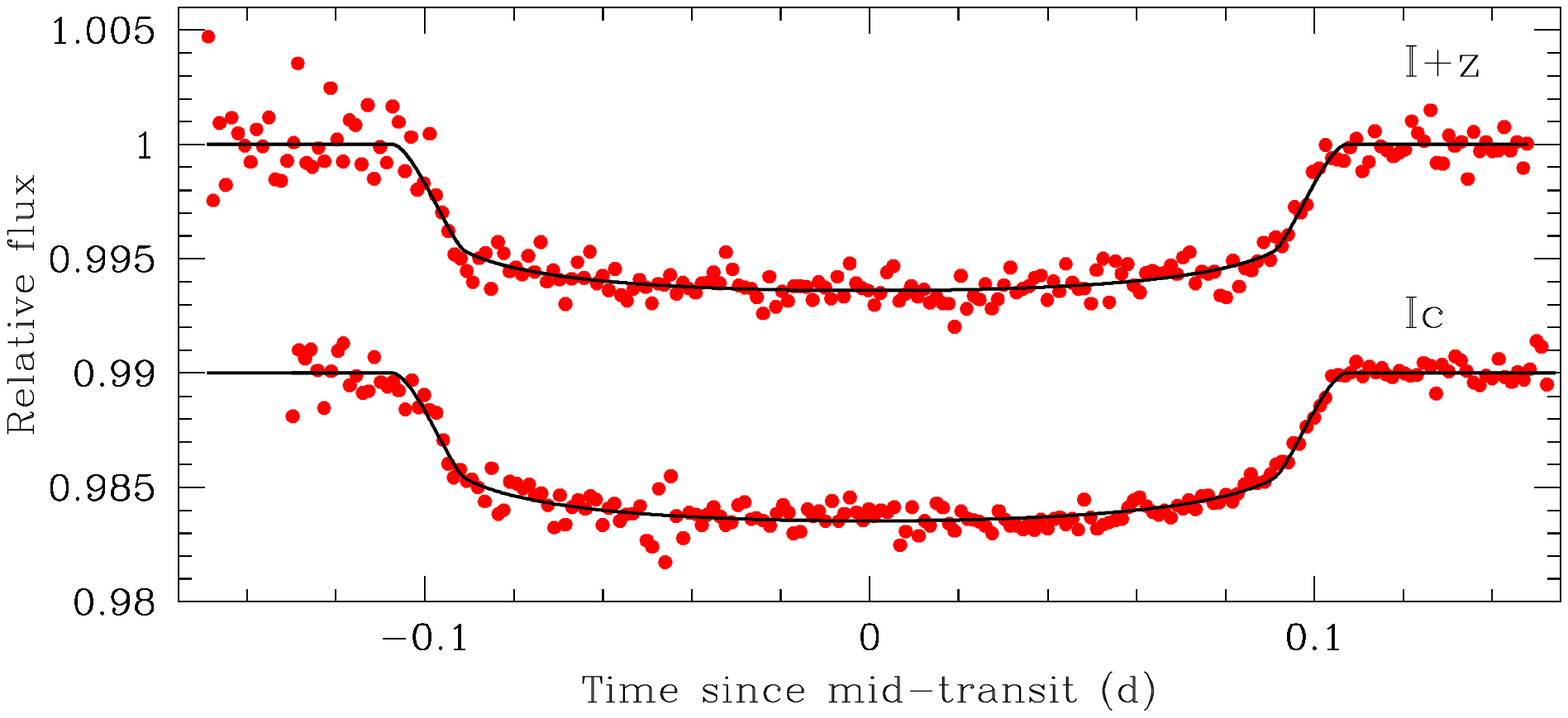}    
\vspace{-5cm}       
\label{lcs68}
\end{figure}

\begin{figure}[h!]
\caption{\textit{Top:} Individual follow-up transit light curves for \hbox{WASP-73 b}. \textit{Bottom:} Combined follow-up photometry for \hbox{WASP-73 b}. The observations are shown as red points (bin width=2min) and are period-folded on the best-fit transit ephemeris. Each light curve has been divided by the respective photometric baseline model (see Section \ref{mcmc}). For each filter, the superimposed, solid, black line is our best-fit transit model. The light curves are shifted along the \textit{y}-axis for clarity.}\label{lcs73}
\centering   
\includegraphics[angle=0, width=0.48\textwidth, height=10cm]{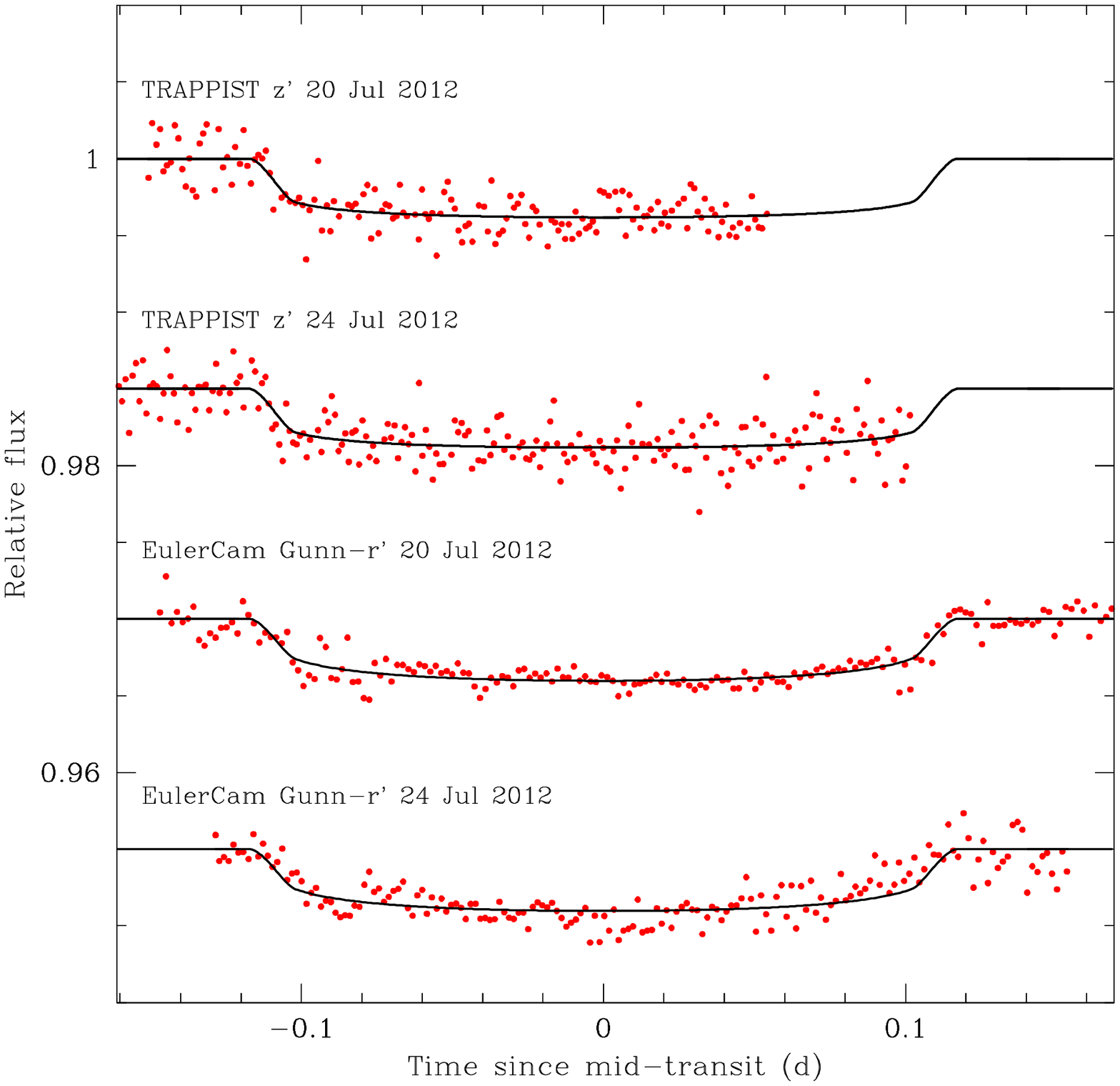}       
\includegraphics[angle=0, width=0.48\textwidth, height=10cm]{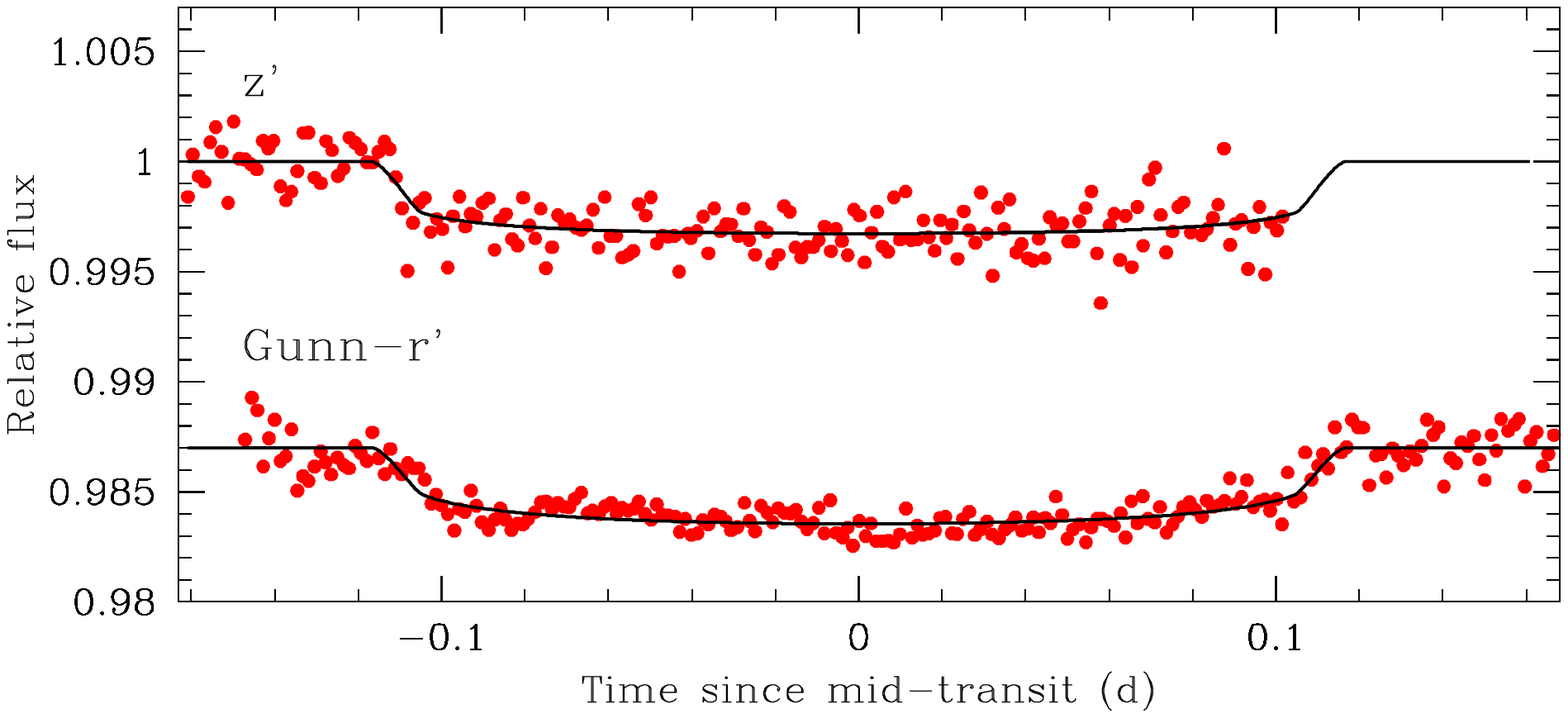}       
\vspace{-5cm}    
\end{figure}

\begin{figure}[h!]  
\caption{\textit{Top:} Individual follow-up transit light curves for \hbox{WASP-88 b}. \textit{Bottom:} Combined follow-up photometry for \hbox{WASP-88 b}. The observations are shown as red points (bin width=2min) and are period-folded on the best-fit transit ephemeris. Each light curve has been divided by the respective photometric baseline model (see Section \ref{mcmc}). For each filter, the superimposed, solid, black line is our best-fit transit model. The light curves are shifted along the \textit{y}-axis for clarity.}
\vspace{0.2cm}
\centering
\includegraphics[angle=0, width=0.48\textwidth, height=10cm]{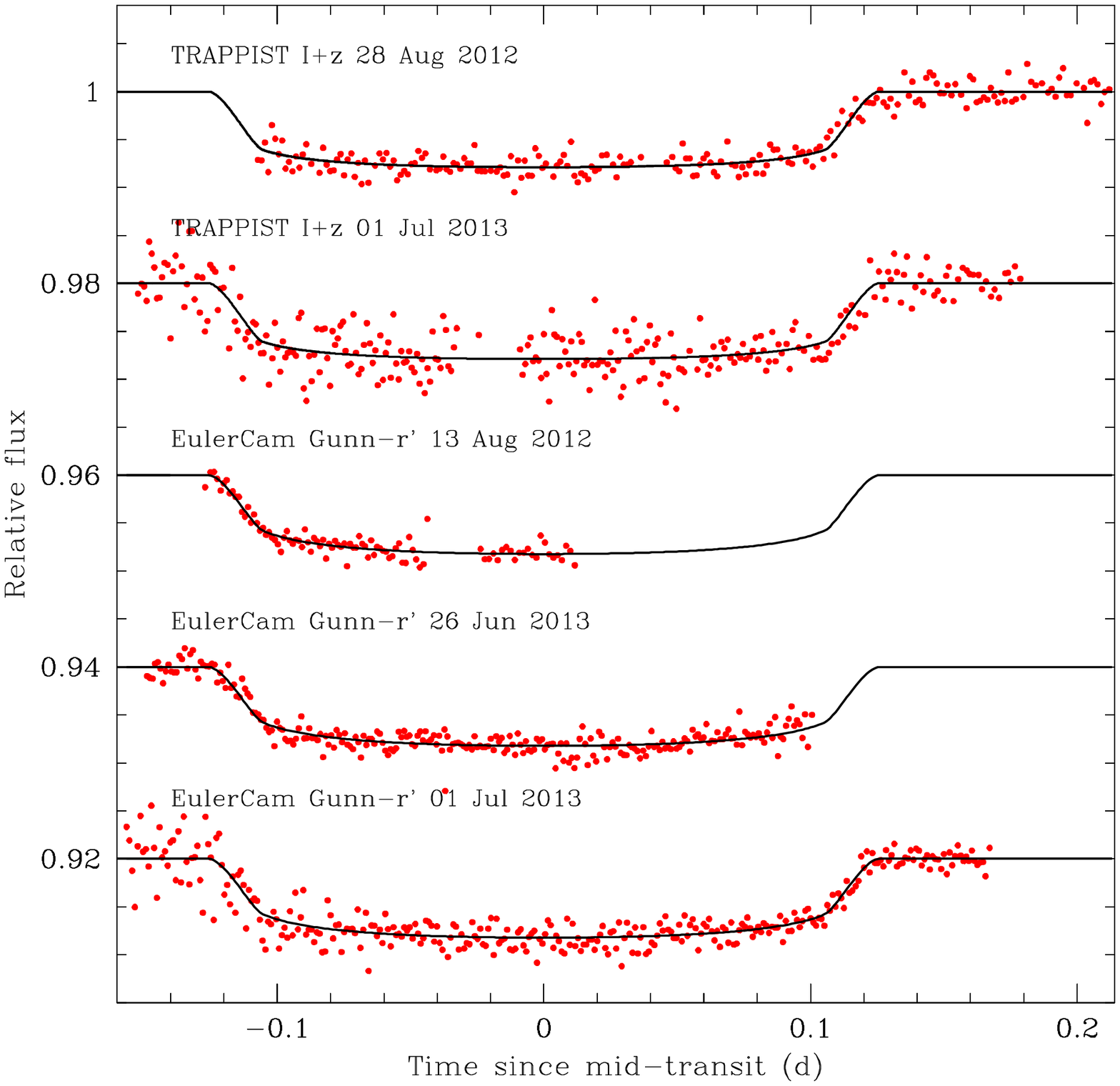} 
\includegraphics[angle=0, width=0.48\textwidth, height=10cm]{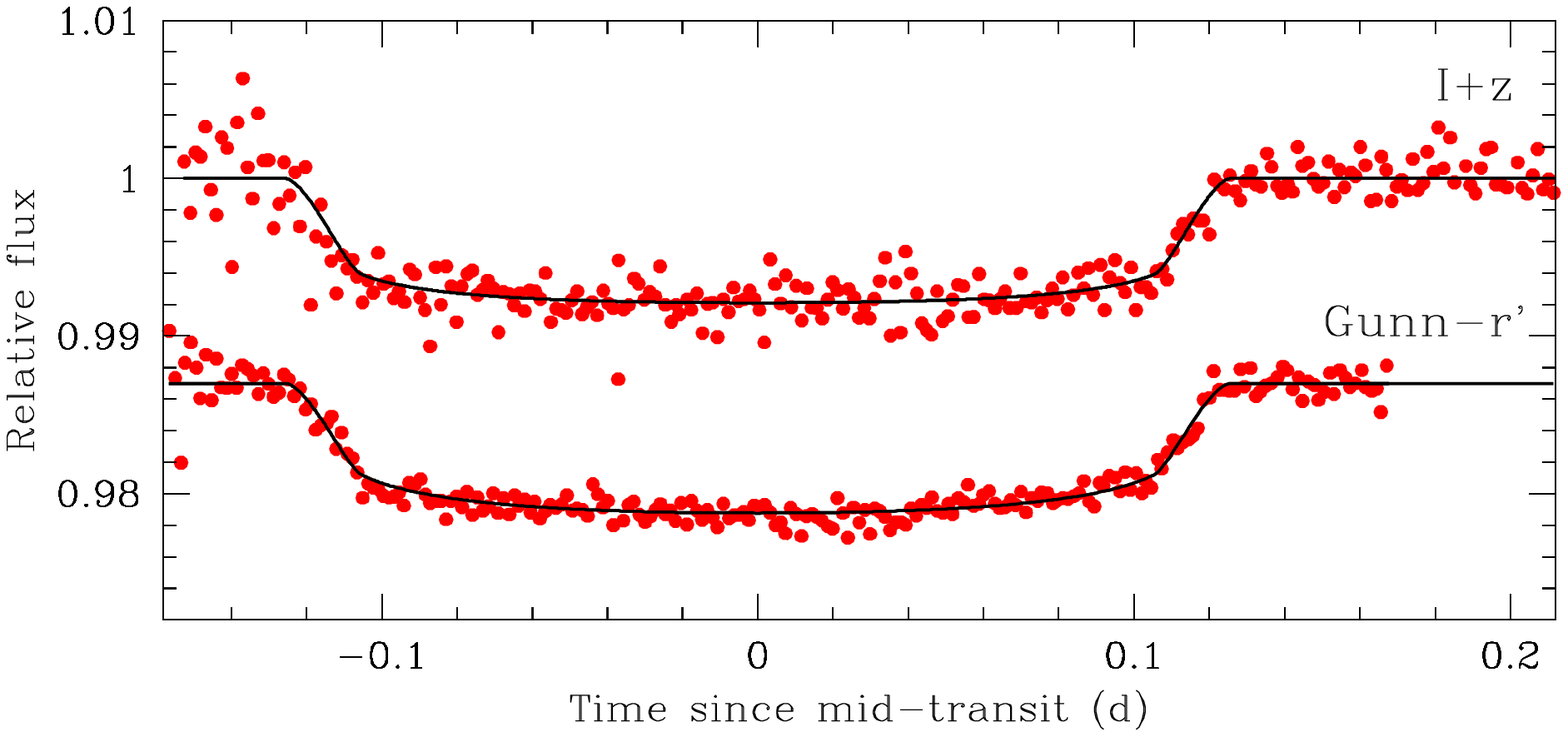}         
\vspace{-5.0cm}    
\label{lcs88}
\end{figure}

\subsection{Global analysis}
\label{mcmc}

\indent
In order to determine the parameters of each system, we performed a combined analysis of the follow-up photometry and the RV data, using for this purpose the adaptive Markov Chain Monte-Carlo (MCMC) code described in \cite{gillon2} and references therein. The algorithm simultaneously models the data, using for the photometry, the transit model by \cite{mandel} multiplied by a different baseline model for each light curve, and for the RVs, a classical Keplerian model (e.g. \citealt{murray}).\\ 
\indent
The photometric baseline models aim to represent astrophysical, instrumental or environmental effects which are able to produce photometric variations and can therefore affect the transit light curves. They are made up of different first to fourth-order polynomials with respect to time or other variables such as airmass, PSF full-width at half maximum, background or stellar position on the detector. In order to find the optimal baseline function for each light curve, i.e. the model minimizing the number of parameters and the level of noise in the best-fit residuals, the Bayes factor, estimated from the Bayesian Information Criterion (\citealt{schwarz}), was used. The best photometric model functions are listed in Table \ref{obstable}. For six TRAPPIST light curves (see Table \ref{obstable}), a normalization offset was also part of the baseline model to represent the effect of the meridian flip, i.e. the $180^{\circ}$ rotation that the German equatorial mount telescope has to undergo when the meridian is reached. This movement results in different positions of the stellar images on the detector before and after the flip and the normalization offset allows to take into account a possible consecutive jump in the differential photometry at the time of the flip.\\ 
\indent
Concerning the RVs, a model with a trend was tested for each system. Such a trend would be indicative of the presence of an additional massive body in the system. A model with a slope was slightly favored in the case of WASP-68, with a Bayes factor $\sim$90. We thus adopted this model for this system but the Bayes factor value is not high enough to be decisive (\citealt{jeffreys}) and more RVs will be needed to confirm this possible trend.\\
\indent
The basic jump parameters in our MCMC analyses, i.e. the parameters that are randomly perturbed at each step of the MCMC, were: the planet/star area ratio $(R_{\mathrm{p}}/R_{\star})^{2}$, the transit impact parameter in case of a circular orbit $b' = a\:\:\mathrm{cos}\:\:\:\:i_{\mathrm{p}}/R_{\star}$ where $a$ is the orbital semi-major axis and $i_{\mathrm{p}}$ is the orbital inclination, the transit width (from 1st to 4th contact) $W$, the time of mid-transit $T_{0}$, the orbital period $P$, the parameter $K_{2} = K \sqrt{1-e^{2}}\:\:P^{1/3}$ where $K$ is the RV orbital semi-amplitude and $e$ is the orbital eccentricity, and the two parameters $\sqrt{e}$ cos $\omega$ and $\sqrt{e}$ sin $\omega$, where $\omega$ is the argument of the periastron. The reasons to use \hbox{$\sqrt{e}$ cos $\omega$} and \hbox{$\sqrt{e}$ sin $\omega$} as jump parameters instead of the more traditional \hbox{$e$ cos $\omega$} and \hbox{$e$ sin $\omega$} are detailed in \cite{triaud3}. For all these jump parameters, we assumed a uniform prior distribution. The photometric baseline model parameters were not actual jump parameters; they were determined by least-square minimization at each step of the MCMC.\\
\indent
The effect of stellar limb-darkening on our transit light curves was accounted for using a quadratic limb-darkening law where the quadratic coefficients $u_{1}$ and $u_{2}$ were allowed to float in our MCMC analysis. However, we used not these coefficients themselves but their combinations $c_{1} = 2 \times u_{1} + u_{2}$ and $c_{2}=u_{1} - 2 \times u_{2}$ as jump parameters, to minimize the correlation of the obtained uncertainties as introduced by \cite{holman}. To obtain a limb-darkening solution consistent with theory, we used normal prior distributions for $u_{1}$ and $u_{2}$ based on theoretical values and 1-$\sigma$ errors interpolated in the tables by \cite{claret}. For the non-standard $I+z$ filter, the modes of the normal prior distributions for $u_{1}$ and $u_{2}$ were taken as the averages of the values interpolated in the tables for the standard filters $I_{c}$ and $z'$, while the errors were computed as the quadratic sums of the errors for these two filters. For the three systems, the prior distributions used for $u_{1}$ and $u_{2}$ are presented in Table 5.

\indent
For each system, a preliminary analysis was performed to determine the correction factors ($CF$) for our photometric errors, as described in \cite{gillon2}. For each light curve, $CF$ is the product of two contributions, $\beta_{w}$ and $\beta_{r}$. On one side, $\beta_{w}$ represents the under- or overestimation of the white noise of each measurement. It is computed as the ratio between the standard deviation of the residuals and the mean photometric error. On the other side, $\beta_{r}$ allows to take into account the correlated noise present in the light curve (i.e. the inability of our model to perfectly fit the data). It is calculated from the standard deviations of the binned and unbinned residuals for different binning intervals ranging from 5 to 120 min, the largest value being kept as $\beta_{r}$. The standard deviation of the best-fit residuals (unbinned and binned per intervals of 2 min), and the deduced values for $\beta_{w}$, $\beta_{r}$ and $CF=\beta_{w} \times \beta_{r}$ for each light curve are presented in Table \ref{obstable}. For each RV time-series, a ``jitter'' noise was determined and added quadratically to the errors in order to equal their mean value to the standard deviation of the best-fit residuals. These RV jitters take into account the instrumental and astrophysical effects (such as stellar activity) that are not included in the initial error estimation. The derived jitter values were 6.6 $\mathrm{m}\:\mathrm{s}^{-1}$ for WASP-68, 9.2 $\mathrm{m}\:\mathrm{s}^{-1}$ for WASP-73 and 10.9 $\mathrm{m}\:\mathrm{s}^{-1}$ for WASP-88.\\
\indent
Our final analyses consisted each of five Markov chains of $10^{5}$ steps, whose convergence was checked using the statistical test of \cite{GR}. At each step of the Markov chains, the stellar density $\rho_{\star}$ was derived from the Kepler's third law and the jump parameters $(R_{\mathrm{p}}/R_{\star})^{2}$, $b'$, $W$, $P$, \hbox{$\sqrt{e}$ cos $\omega$} and \hbox{$\sqrt{e}$ sin $\omega$} (see e.g. \citealt{seager} and \citealt{winn}). The resulting stellar density and values for $T_{\mathrm{eff}}$ and $[$Fe/H$]$ drawn from the normal distributions deduced from our spectroscopic analysis (see Section \ref{barry}) were used to determine a value for the stellar mass $M_{\star}$ through an empirical law \hbox{$M_{\star}$($\rho_{\star}$, $T_{\mathrm{eff}}$, $[$Fe/H$]$)} (\citealt{enoch}, \citealt{gillon3}) calibrated using the set of well-constrained detached binary systems presented by \cite{southworth}. For WASP-68, this set was reduced to the 116 stars with a mass between 0.7 and \hbox{1.7 $M_{\odot}$}, while the 119 stars with a mass between 0.9 and \hbox{1.9 $M_{\odot}$} were used for WASP-73 and WASP-88. The goal of these selections was to benefit from our preliminary estimation of the stellar masses (see Section \ref{barry}) to improve the determination of the \hbox{systems' parameters}. In order to propagate correctly the error on the empirical law, the parameters of the selected subset of calibration stars were normally perturbed within their observational error bars and the coefficients of the law were redetermined at each MCMC step. The other physical parameters were then deduced from the jump parameters and stellar mass.\\
\indent
For each system, two analyses were performed : one assuming a circular orbit ($e$ = 0) and one with a free eccentricity. For the three systems, the resulting Bayes factors ($\sim$1100 for \hbox{WASP-68}, $\sim$1800 for \hbox{WASP-73} and \hbox{WASP-88}) favored the circular solutions. We thus adopt the corresponding results as our nominal solutions but for the sake of completeness, we present the derived parameters for both models in Tables \ref{w68table} (WASP-68), \ref{w73table} (WASP-73) and \ref{w88table} (WASP-88). The best-fit transit models for the circular solutions are shown in Figures \ref{lcs68}, \ref{lcs73} and \ref{lcs88}.

\begin{table}[h!]
\centering
\begin{tabular}{lccc}
  \hline
  \hline
  \textbf{LD coefficient} & \textbf{WASP-68} & \textbf{WASP-73} & \textbf{WASP-88}\\
  \hline
  $u_{1,I+z}$&0.256 $\pm$ 0.021&-&0.187 $\pm$ 0.016\\
  $u_{2,I+z}$&0.283 $\pm$ 0.003&-&0.303 $\pm$ 0.005\\
  $u_{1,I_{c}}$&0.275 $\pm$ 0.010&-&-\\
  $u_{2,I_{c}}$&0.286 $\pm$ 0.005&-&-\\
  $u_{1,z'}$&-& 0.213 $\pm$ 0.015&-\\
  $u_{2,z'}$&-& 0.291 $\pm$ 0.005&-\\
  $u_{1,\mathrm{Gunn}-r'}$&-&0.349 $\pm$ 0.020&0.293 $\pm$ 0.012\\
  $u_{2,\mathrm{Gunn}-r'}$&-&0.301 $\pm$ 0.008&0.319 $\pm$ 0.005\\
  \hline
  \hline
\end{tabular}
\label{ldtable}
\caption{Expectations and standard deviations of the normal distributions used as prior distributions for the quadratic limb-darkening (LD) coefficients $u_{1}$ and $u_{2}$ in our MCMC analyses.}
\vspace{-0.9cm}
\end{table}


\subsection{Stellar evolution modeling}
\label{val}

\begin{figure}[h!]
\includegraphics[height=8cm,width=8.5cm]{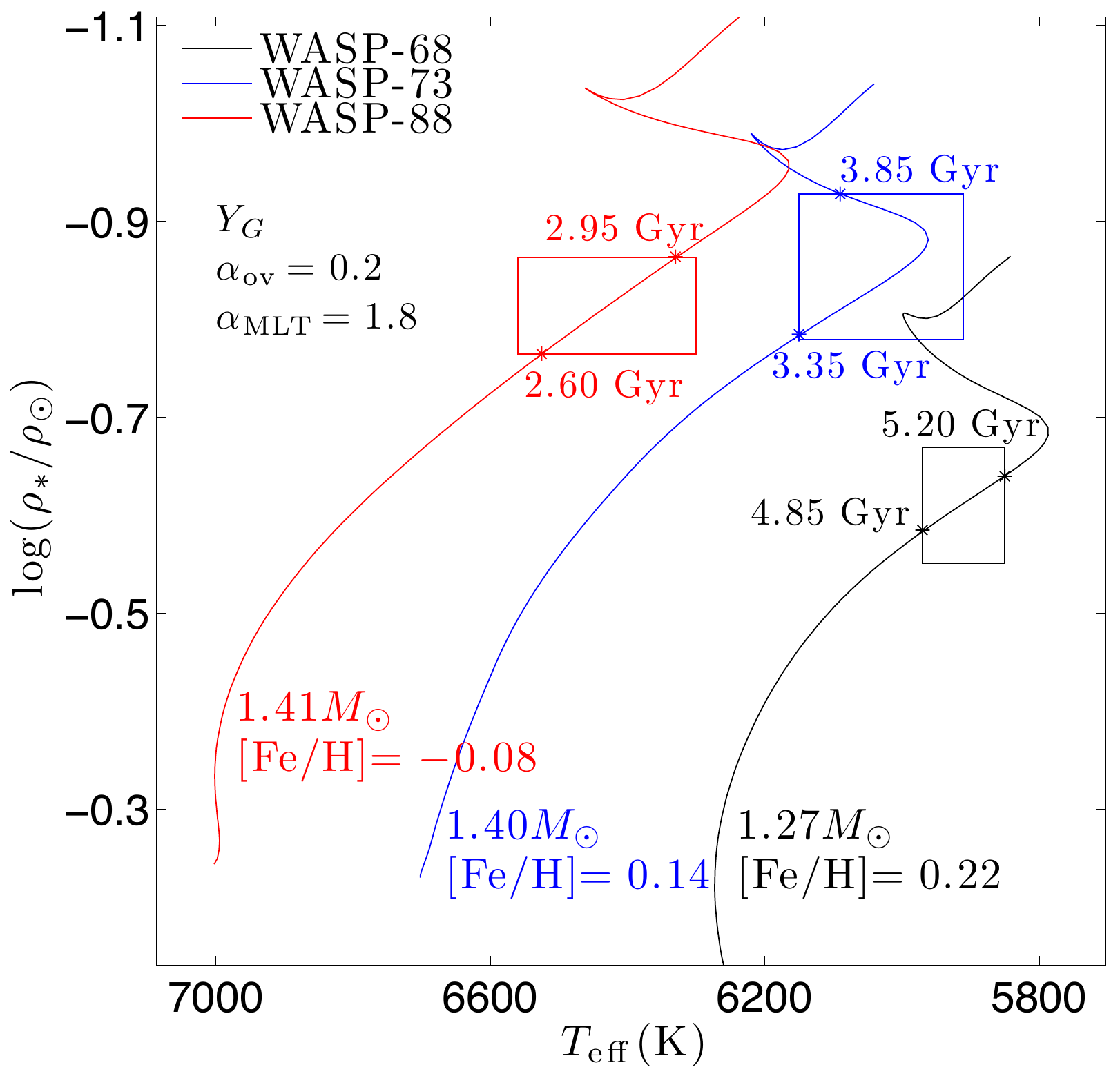}    
\caption{Evolutionary tracks in a $T_{\rm eff} - \log(\rho_*/\rho_{\odot})$ HR diagram for WASP-68 (black), WASP-73 (blue), and WASP-88 (red), for their respective central masses and metallicities. These evolutionary tracks span from the zero-age main sequence to the beginning of the subgiant phase. The ages of the stars when they cross their respective 1-$\sigma$ box $T_{\rm eff} - \log(\rho_*/\rho_{\odot})$ are also indicated.}
\label{hr}
\end{figure}

After the completion of the MCMC analyses, we performed for the three systems a stellar evolution modeling based on the CLES code (\citealt{scuflaire}), with the aim to assess the reliability of the deduced stellar masses and to estimate the age of the systems. We used as inputs the stellar densities deduced from the MCMC analyses, and the effective temperatures and metallicities as derived from spectroscopy (see Tables 6, 7, and 8). We considered here that [Fe/H] represents the global metallicity with respect to the Sun, defined as $[\log (Z/X)_* - \log (Z/X)_{\odot}]$, where $X$ and $Z$ are the fractional mass of hydrogen and elements heavier than helium respectively. We used the most recent solar mixture of \cite{asplund}, giving for the current Sun $(Z/X)_{\odot} = 0.0181$. The parameter of the mixing-length theory (MLT) of convection was kept fixed to the solar calibration ($\alpha_{\rm MLT} = 1.8$), and microscopic diffusion (gravitational settling) of elements was included.\\
\indent
The resulting stellar masses are \hbox{1.27 $\pm$ 0.11 $M_{\odot}$} \hbox{(WASP-68)}, \hbox{1.40 $\pm$ 0.16 $M_{\odot}$} \hbox{(WASP-73)}, and \hbox{1.41 $\pm$ 0.14 $M_{\odot}$} \hbox{(WASP-88)}. These 1-$\sigma$ uncertainties were obtained by considering the respective 1-$\sigma$ range for the effective temperatures, metallicities and stellar densities, but also by varying the internal stellar physics. We indeed computed, since the helium atmospheric abundance cannot be directly measured from spectroscopy, evolutionary tracks with two initial helium abundances: the solar value ($Y_{\odot}=0.2485$), and a value labelled $Y_G$ that increases with $Z$ (as expected if the local medium follows the general trend observed for the chemical evolution of galaxies; \citealt{izotov}). We also investigated the effects of the possible convective core overshooting, by varying $\alpha_{\mathrm{ov}}$ between 0 and 0.3. Within the same hypotheses, the resulting stellar ages range \hbox{4.2$-$8.3 Gyr} (\hbox{WASP-68}), \hbox{2.7$-$6.4 Gyr} (\hbox{WASP-73}), and \hbox{1.8-5.3 Gyr} (\hbox{WASP-88}). Three evolutionary tracks, respectively for the central value for the stellar mass and metallicity of WASP-68, WASP-73, and WASP-88, are shown on Fig. \ref{hr}. These evolutionary tracks span from the beginning (zero-age) of the main sequence to the beginning of the subgiant phase (core H-burning exhaustion). WASP-73 appears to be the most evolved star,  close to or already in the subgiant phase. WASP-68 and WASP-88 are less evolved, although in an advanced stage of core H-burning. The subgiant phase is also a possibility, although very unlikely.\\
\indent
The masses derived for WASP-68 and WASP-88 are in excellent agreement with the MCMC results obtained through an empirical law \hbox{$M_{\star}$($\rho_{\star}$, $T_{\mathrm{eff}}$, $[$Fe/H$]$)} calibrated using a set of well-constrained detached eclipsing binary (EB) systems (see Section 3.2). The agreement is also good for WASP-73, which is close to core H-burning exhaustion or already in the subgiant phase, despite that the EB sample contains only a small fraction of significantly evolved objects. This shows that the EB empirical law used in the MCMC analyses is valid for the 3 stars considered here.\\ 
\indent
For even more evolved stars, the EB empirical law would reach its limit of applicability and could lead to inaccurate results. In such a case, a more reliable alternative would be to implement the stellar evolutionary models in the MCMC analysis, by assuming realistic prior probability distributions on the different stellar physics parameters (overshooting, diffusion, initial composition, etc.) and computing at each step $M_*$ from $\rho_*$, $T_{\rm eff}$, and [Fe/H]. This is a long-term goal we are pursuing (e.g. \citealt{triaud3}). Obtaining an accurate stellar mass from evolution modeling primarily needs accurate spectroscopic estimates for the effective temperature but also, very importantly, for the metallicity (compare in \hbox{Fig. \ref{hr}} the tracks of two very close stellar masses, \hbox{1.40 $M_{\odot}$} and \hbox{1.41 $M_{\odot}$}, but with quite different metallicities). 

\begin{table*}[t!]
\centering
\caption{\textbf{System parameters for WASP-68}. The values given for the parameters derived from our MCMC analyses are medians and 1-$\sigma$ limits of the marginalized posterior probability distributions. $ ^{a}$Using as priors the values derived from the spectroscopic analysis. $ ^{b}$Using $a_{\mathrm{R}}$ = 2.46 $R_{\mathrm{p}}(M_{\star}/M_{\mathrm{p}})^{1/3}$ (\citealt{chandr}). $ ^{c}$Assuming a null Bond albedo and an efficient heat distribution between both hemispheres.}
\begin{tabular}{lcc}
  \hline
  \hline
   & \textbf{General information} & \\
  \hline
  \hline
  RA (J2000) & 20 20 22.98 & \\
  Dec (J2000) & -19 18 52.9 & \\
  $V$ & 10.7 & \\
  $K$ & 8.9 & \\
  \hline
  \hline
   & \textbf{Stellar parameters from spectroscopic analysis} & \\
  \hline
  \hline
  $T_{\mathrm{eff}}$ (K) & 5910 $\pm$ 60 & \\
  log $g_{\star}$  [cgs] & 4.17 $\pm$ 0.11 & \\
  $[$Fe/H$]$ & 0.22 $\pm$ 0.08 & \\
  $\xi_{\mathrm{t}}$ (km $\mathrm{s}^{-1}$) & 1.4 $\pm$ 0.1 & \\
  $v_{\mathrm{mac}}$ (km $\mathrm{s}^{-1}$) & 2.6 $\pm$ 0.3 & \\
  $v$ sin $i_{\star}$ (km $\mathrm{s}^{-1}$) & 2.3 $\pm$ 0.8 & \\
  Sp. type & G0 & \\
  \hline
  \hline
   & \textbf{Parameters from MCMC analyses} & \\
  \hline
  \hline
  \textbf{Jump parameters} & $e \ge$ 0 & \textbf{$e$ = 0 (adopted)} \\
  \hline
  Planet/star area ratio $(R_{\mathrm{p}}/R_{\star})^{2}$ [$\%$] & 0.57$\pm$0.03 & $0.57_{-0.02}^{+0.03}$\\
  $b' = a\:\mathrm{cos}\:i_{\mathrm{p}}/R_{\star}\:\: [R_{\star}]$ & $0.27_{-0.18}^{+0.16}$ & $0.26_{-0.18}^{+0.15}$\\
  Transit width $W$ [d] & $0.214_{-0.002}^{+0.003}$ & $0.214_{-0.002}^{+0.003}$\\
  $T_{0}$ [$\mathrm{HJD_{TDB}}$] & $2456064.86355_{-0.00062}^{+0.00064}$ & $2456064.86356_{-0.00061}^{+0.00060}$ \\ 
  Orbital period $P$ [d] & 5.084299 $\pm$ 0.000015 & 5.084298 $\pm$ 0.000015\\
  RV $K_{2}$ [$\mathrm{m\:s^{-1}\:d^{1/3}}$] & 168.2 $\pm$ 3.3 & $168.3_{-3.3}^{+3.2}$\\
  RV slope [$\mathrm{m\:s^{-1}\:y^{-1}}$] & 14 $\pm$ 2 & 14 $\pm$ 2\\  
  $\sqrt{e}$ cos $\omega$ & $0.091_{-0.058}^{+0.041}$ & 0 (fixed)\\
  $\sqrt{e}$ sin $\omega$ & $-0.037_{-0.091}^{+0.100}$ & 0 (fixed)\\
  $c_{1,I+z}$ & 0.79$\pm$0.04 & $0.79_{-0.04}^{+0.05}$\\
  $c_{2,I+z}$ & -0.31 $\pm$ 0.02 & -0.31$\pm$0.03\\
  $c_{1,I_{c}}$ & 0.84 $\pm$ 0.03 & 0.84 $\pm$ 0.02\\
  $c_{2,I_{c}}$ & -0.29 $\pm$ 0.02 & -0.29 $\pm$ 0.02\\
  Effective temperature $T_{\mathrm{eff}}$ $\mathrm{[K]}^{a}$ & 5911 $\pm$ 60 & $5911_{-60}^{+59}$\\
  Metallicity [Fe/H] $\mathrm{[dex]}^{a}$ & 0.22 $\pm$ 0.08 & 0.22 $\pm$ 0.08\\
  \hline
  \textbf{Deduced stellar parameters} & & \\
  \hline
  Mean density $\rho_{\star}$ [$\rho_{\odot}$] & $0.26_{-0.05}^{+0.03}$ & $0.26_{-0.05}^{+0.03}$\\
  Surface gravity log $g_{\star}$ [cgs] & $4.09_{-0.08}^{+0.13}$ & $4.09_{-0.08}^{+0.13}$\\
  Mass $M_{\star}$ [$M_{\odot}$] & 1.23 $\pm$ 0.03 & 1.24 $\pm$ 0.03\\
  Radius $R_{\star}$ [$R_{\odot}$] & $1.69_{-0.07}^{+0.13}$ & $1.69_{-0.06}^{+0.11}$\\
  Luminosity $L_{\star}$ [$L_{\odot}$] & $3.1_{-0.3}^{+0.5}$ & $3.2_{-0.3}^{+0.4}$\\
  $u_{1,I+z}$ & 0.25 $\pm$ 0.02 & $0.26_{-0.02}^{+0.03}$\\
  $u_{2,I+z}$ & 0.28 $\pm$ 0.01 & 0.28 $\pm$ 0.01\\
  $u_{1,I_{c}}$ & 0.28 $\pm$ 0.01 & 0.28 $\pm$ 0.01\\
  $u_{2,I_{c}}$ & 0.28 $\pm$ 0.01 & 0.28 $\pm$ 0.01\\
  \hline
  \textbf{Deduced planet parameters} & & \\
  \hline
  RV $K$ [$\mathrm{m\:s^{-1}}$] & 97.9 $\pm$ 1.9 & 97.9 $\pm$ 1.9\\
  Planet/star radius ratio $R_{\mathrm{p}}/R_{\star}$ & 0.075 $\pm$ 0.002 & 0.075 $\pm$ 0.002\\
  $b_{\mathrm{tr}}$ [$R_{\star}$] & $0.27_{-0.18}^{+0.16}$ & $0.26_{-0.18}^{+0.15}$\\
  $b_{\mathrm{oc}}$ [$R_{\star}$] & $0.27_{-0.18}^{+0.16}$ & $0.26_{-0.18}^{+0.15}$\\
  $T_{\mathrm{oc}}$ [$\mathrm{HJD_{TDB}}$] & 2456067.444 $\pm$ 0.029 & 2456067.406 $\pm$ 0.001\\
  Scaled semi-major axis $a/R_{\star}$ & $7.91_{-0.50}^{+0.29}$ & $7.90_{-0.46}^{+0.25}$\\
  Orbital semi-major axis $a$ [AU] & $0.06204_{-0.00042}^{+0.00049}$ & $0.06206_{-0.00040}^{+0.00045}$\\
  Orbital inclination $i_{\mathrm{p}}$ [deg] & $88.1_{-1.4}^{+1.3}$ & 88.1 $\pm$ 1.3\\
  Orbital eccentricity $e$ & $0.017_{-0.010}^{+0.012}$, $<$ 0.063 (95 $\%$) & 0 (fixed)\\
  Argument of periastron $\omega$ [deg] & $338_{-42}^{+63}$ & - \\
  Mean density $\rho_{\mathrm{p}}$ [$\rho_{\mathrm{Jup}}$] & $0.50_{-0.11}^{+0.08}$ & $0.50_{-0.10}^{+0.07}$\\
  Surface gravity log $g_{\mathrm{p}}$ [cgs] & $3.19_{-0.07}^{+0.04}$ & $3.19_{-0.06}^{+0.04}$\\
  Mass $M_{\mathrm{p}}$ [$M_{\mathrm{Jup}}$] & 0.95 $\pm$ 0.03 & 0.95 $\pm$ 0.03\\
  Radius $R_{\mathrm{p}}$ [$R_{\mathrm{Jup}}$] & $1.27_{-0.06}^{+0.11}$ & $1.24_{-0.06}^{+0.10}$\\ 
  Roche limit $a_{\mathrm{R}}$ $\mathrm{[AU]}^{b}$ & $0.01413_{-0.00075}^{+0.00130}$ & $0.01415_{-0.00065}^{+0.00120}$\\
  $a/a_{\mathrm{R}}$ & $4.39_{-0.35}^{+0.23}$ & $4.38_{-0.32}^{+0.19}$\\
  Equilibrium temperature $T_{\mathrm{eq}}$ $\mathrm{[K]}^{c}$ & $1488_{-32}^{+49}$ & $1490_{-29}^{+44}$\\
  Irradiation [erg $\mathrm{s}^{-1} \mathrm{cm}^{-2}$] & $1.1^{+0.3}_{-0.2}\:10^{9}$ & $1.1^{+0.3}_{-0.2}\:10^{9}$\\
  \hline
  \hline
\end{tabular}
\label{w68table}
\end{table*}

\begin{table*}[t!]
\centering
\caption{\textbf{System parameters for WASP-73}. The values given for the parameters derived from our MCMC analyses are medians and 1-$\sigma$ limits of the marginalized posterior probability distributions. $ ^{a}$Using as priors the values derived from the spectroscopic analysis. $ ^{b}$Using $a_{\mathrm{R}}$ = 2.46 $R_{\mathrm{p}}(M_{\star}/M_{\mathrm{p}})^{1/3}$ (\citealt{chandr}). $ ^{c}$Assuming a null Bond albedo and an efficient heat distribution between both hemispheres.}
\begin{tabular}{lcc}
  \hline
  \hline
   & \textbf{General information} & \\
  \hline
  \hline
  RA (J2000) & 21 19 47.91 & \\
  Dec (J2000) & -58 08 56.0 &\\
  $V$ & 10.5 & \\
  $K$ & 9.0 & \\
  \hline
  \hline
   & \textbf{Stellar parameters from spectroscopic analysis} & \\
  \hline
  \hline
  $T_{\mathrm{eff}}$ (K) & 6030 $\pm$ 120 & \\
  log $g_{\star}$ & 3.92 $\pm$ 0.08 & \\
  $[$Fe/H$]$ & 0.14 $\pm$ 0.14 & \\
  $\xi_{\mathrm{t}}$ (km $\mathrm{s}^{-1}$) & 1.1 $\pm$ 0.2 & \\
  $v_{\mathrm{mac}}$ (km $\mathrm{s}^{-1}$) & 3.3 $\pm$ 0.3 & \\
  $v$ sin $i_{\star}$ (km $\mathrm{s}^{-1}$) & 6.1 $\pm$ 0.6 & \\
  Sp. type & F9 & \\
  \hline
  \hline
   & \textbf{Parameters from MCMC analyses} & \\
  \hline
  \hline
  \textbf{Jump parameters} & $e \ge$ 0 & \textbf{$e$ = 0 (adopted)} \\
  \hline
  Planet/star area ratio $(R_{\mathrm{p}}/R_{\star})^{2}$ [$\%$] & 0.33 $\pm$ 0.03 & 0.33 $\pm$ 0.03\\
  $b' = a\:\mathrm{cos}\:i_{\mathrm{p}}/R_{\star}\:\: [R_{\star}]$ & $0.26_{-0.17}^{+0.20}$ & $0.26_{-0.18}^{+0.20}$\\
  Transit width $W$ [d] & 0.233 $\pm$ 0.003 & 0.233 $\pm$ 0.003\\
  $T_{0}$ [$\mathrm{HJD_{TDB}}$] & 2456128.7063 $\pm$ 0.0011 & 2456128.7063 $\pm$ 0.0011\\ 
  Orbital period $P$ [d] & 4.08721 $\pm$ 0.00022 & 4.08722 $\pm$ 0.00022\\
  RV $K_{2}$ [$\mathrm{m\:s^{-1}\:d^{1/3}}$] & 313.5 $\pm$ 6.9 & 313.9 $\pm$ 6.6\\
  $\sqrt{e}$ cos $\omega$ & $-0.021_{-0.061}^{+0.065}$ & 0 (fixed)\\
  $\sqrt{e}$ sin $\omega$ & $0.039_{-0.110}^{+0.100}$ & 0 (fixed)\\
  $c_{1,z'}$ & 0.71 $\pm$ 0.03 & $0.71_{-0.03}^{+0.04}$\\
  $c_{2,z'}$ & -0.37 $\pm$ 0.02 & -0.37 $\pm$ 0.02\\
  $c_{1,\mathrm{Gunn}-r'}$ & $1.01_{-0.05}^{+0.04}$ & $1.01_{-0.05}^{+0.04}$\\
  $c_{2,\mathrm{Gunn}-r'}$ & -0.25 $\pm$ 0.03 & -0.25 $\pm$ 0.03\\
  Effective temperature $T_{\mathrm{eff}}$ $\mathrm{[K]}^{a}$ & 6030 $\pm$ 120 & 6036 $\pm$ 120\\
  Metallicity [Fe/H] $\mathrm{[dex]}^{a}$ & 0.14 $\pm$ 0.14 & 0.14 $\pm$ 0.14\\
  \hline
  \textbf{Deduced stellar parameters} & & \\
  \hline
  Mean density $\rho_{\star}$ [$\rho_{\odot}$] & $0.15_{-0.03}^{+0.02}$ & $0.15_{-0.04}^{+0.02}$\\
  Surface gravity log $g_{\star}$ [cgs] & $3.92_{-0.06}^{+0.04}$ & $3.93_{-0.06}^{+0.04}$\\
  Mass $M_{\star}$ [$M_{\odot}$] & $1.34_{-0.04}^{+0.05}$ & $1.34_{-0.04}^{+0.05}$\\
  Radius $R_{\star}$ [$R_{\odot}$] & $2.09_{-0.09}^{+0.18}$ & $2.07_{-0.08}^{+0.19}$\\
  Luminosity $L_{\star}$ [$L_{\odot}$] & $5.2_{-0.7}^{+1.0}$ & $5.2_{-0.6}^{+1.0}$\\
  $u_{1,z'}$ & 0.21 $\pm$ 0.02 & 0.21 $\pm$ 0.02\\
  $u_{2,z'}$ & 0.29 $\pm$ 0.01 & 0.29 $\pm$ 0.01\\
  $u_{1,\mathrm{Gunn}-r'}$ & 0.35 $\pm$ 0.03 & 0.35 $\pm$ 0.03\\
  $u_{2,\mathrm{Gunn}-r'}$ & 0.30 $\pm$ 0.01 & 0.30 $\pm$ 0.01\\
  \hline
  \textbf{Deduced planet parameters} & & \\
  \hline
  RV $K$ [$\mathrm{m\:s^{-1}}$] & 196.1 $\pm$ 4.3 & 196.3 $\pm$ 4.1\\
  Planet/star radius ratio $R_{\mathrm{p}}/R_{\star}$ & 0.057 $\pm$ 0.003 & 0.057 $\pm$ 0.003\\
  $b_{\mathrm{tr}}$ [$R_{\star}$] & $0.25_{-0.17}^{+0.20}$ & $0.26_{-0.18}^{+0.20}$\\
  $b_{\mathrm{oc}}$ [$R_{\star}$] & $0.26_{-0.17}^{+0.20}$ & $0.26_{-0.18}^{+0.20}$\\
  $T_{\mathrm{oc}}$ [$\mathrm{HJD_{TDB}}$] & $2456130.746_{-0.021}^{+0.013}$ & 2456130.750 $\pm$ 0.002\\
  Scaled semi-major axis $a/R_{\star}$ & $5.68_{-0.42}^{+0.21}$ & $5.73_{-0.45}^{+0.18}$\\
  Orbital semi-major axis $a$ [AU] & $0.05514_{-0.00054}^{+0.00061}$ & $0.05512_{-0.00053}^{+0.00060}$\\
  Orbital inclination $i_{\mathrm{p}}$ [deg] & $87.4_{-2.4}^{+1.8}$ & $87.4_{-2.4}^{+1.8}$\\
  Orbital eccentricity $e$ & $0.011_{-0.008}^{+0.015}$, $<$ 0.074 (95 $\%$) & 0 (fixed)\\
  Argument of periastron $\omega$ [deg] & $108_{-68}^{+110}$ & - \\
  Mean density $\rho_{\mathrm{p}}$ [$\rho_{\mathrm{Jup}}$] & $1.19_{-0.29}^{+0.25}$ & $1.20_{-0.30}^{+0.26}$\\
  Surface gravity log $g_{\mathrm{p}}$ [cgs] & $3.54_{-0.08}^{+0.06}$ & $3.54_{-0.08}^{+0.06}$\\
  Mass $M_{\mathrm{p}}$ [$M_{\mathrm{Jup}}$] & 1.88 $\pm$ 0.06 & $1.88_{-0.06}^{+0.07}$\\
  Radius $R_{\mathrm{p}}$ [$R_{\mathrm{Jup}}$] & $1.16_{-0.08}^{+0.12}$ & $1.16_{-0.08}^{+0.12}$\\
  Roche limit $a_{\mathrm{R}}$ $\mathrm{[AU]}^{b}$ & $0.01090_{-0.00072}^{+0.00120}$ & $0.01089_{-0.00072}^{+0.00120}$\\
  $a/a_{\mathrm{R}}$ & $5.05_{-0.45}^{+0.34}$ & $5.07_{-0.46}^{+0.34}$\\ 
  Equilibrium temperature $T_{\mathrm{eq}}$ $\mathrm{[K]}^{c}$ & $1795_{-52}^{+73}$ & $1790_{-51}^{+75}$\\
  Irradiation [erg $\mathrm{s}^{-1} \mathrm{cm}^{-2}$] & $2.4^{+0.7}_{-0.4}\:10^{9}$ & $2.3^{+0.8}_{-0.4}\:10^{9}$ \\
  \hline
  \hline
\end{tabular}
\vspace{1cm}
\label{w73table}
\end{table*}

\begin{table*}[t!]
\centering
\caption{\textbf{System parameters for WASP-88}. The values given for the parameters derived from our MCMC analyses are medians and 1-$\sigma$ limits of the marginalized posterior probability distributions. $ ^{a}$Using as priors the values derived from the spectroscopic analysis. $ ^{b}$Using $a_{\mathrm{R}}$ = 2.46 $R_{\mathrm{p}}(M_{\star}/M_{\mathrm{p}})^{1/3}$ (\citealt{chandr}). $ ^{c}$Assuming a null Bond albedo and an efficient heat distribution between both hemispheres.}
\begin{tabular}{lcc}
  \hline
  \hline
   & \textbf{General information} & \\
  \hline
  \hline
  RA (J2000) & 20 38 02.70 & \\
  Dec (J2000) & -48 27 43.2 & \\
  $V$ & 11.4 & \\
  $K$ & 10.3 & \\
   \hline
  \hline
   & \textbf{Stellar parameters from spectroscopic analysis} & \\
  \hline
  \hline
  $T_{\mathrm{eff}}$ (K) & 6430 $\pm$ 130 & \\
  log $g_{\star}$ & 4.03 $\pm$ 0.09 & \\
  $[$Fe/H$]$ & -0.08 $\pm$ 0.12 & \\
  $\xi_{\mathrm{t}}$ (km $\mathrm{s}^{-1}$) & 1.4 $\pm$ 0.1 & \\
  $v_{\mathrm{mac}}$ (km $\mathrm{s}^{-1}$) & 4.7 $\pm$ 0.3 & \\
  $v$ sin $i_{\star}$ (km $\mathrm{s}^{-1}$) & 8.4 $\pm$ 0.8 & \\
  Sp. type & F6 & \\
  \hline
  \hline
   & \textbf{Parameters from MCMC analyses} & \\
  \hline
  \hline
  \textbf{Jump parameters} & $e \ge$ 0 & \textbf{$e$ = 0 (adopted)} \\
  \hline
  Planet/star area ratio $(R_{\mathrm{p}}/R_{\star})^{2}$ [$\%$] & $0.71_{-0.03}^{+0.04}$ & 0.70 $\pm$ 0.03\\
  $b' = a\:\mathrm{cos}\:i_{\mathrm{p}}/R_{\star}\:\: [R_{\star}]$ & $0.24_{-0.16}^{+0.15}$ & 0.23 $\pm$ 0.15\\
  Transit width $W$ [d] & $0.252_{-0.002}^{+0.003}$ & $0.252_{-0.002}^{+0.003}$\\
  $T_{0}$ [$\mathrm{HJD_{TDB}}$] & $2456474.73145_{-0.00089}^{+0.00084}$ & $2456474.73154_{-0.00086}^{+0.00085}$\\ 
  Orbital period $P$ [d] & 4.954000 $\pm$ 0.000019 & 4.954000 $\pm$ 0.000019\\
  RV $K_{2}$ [$\mathrm{m\:s^{-1}\:d^{1/3}}$] & 90.3 $\pm$ 11.0 & 89.4 $\pm$ 11.0\\
  $\sqrt{e}$ cos $\omega$ & $-0.147_{-0.140}^{+0.190}$ & 0 (fixed)\\
  $\sqrt{e}$ sin $\omega$ & $-0.010_{-0.250}^{+0.260}$ & 0 (fixed)\\
  $c_{1,I+z}$ & 0.67 $\pm$ 0.04 & 0.67 $\pm$ 0.04\\
  $c_{2,I+z}$ & -0.42 $\pm$ 0.02 & -0.42 $\pm$ 0.02\\
  $c_{1,\mathrm{Gunn}-r'}$ & 0.90 $\pm$ 0.03 & 0.90 $\pm$ 0.03\\
  $c_{2,\mathrm{Gunn}-r'}$ & -0.35 $\pm$ 0.02 & -0.35 $\pm$ 0.02\\
  Effective temperature $T_{\mathrm{eff}}$ $\mathrm{[K]}^{a}$ & 6431 $\pm$ 130 & 6431 $\pm$ 130\\
  Metallicity [Fe/H] $\mathrm{[dex]}^{a}$ & -0.08 $\pm$ 0.12 & -0.08 $\pm$ 0.12\\
  \hline
  \textbf{Deduced stellar parameters} & & \\
  \hline
  Mean density $\rho_{\star}$ [$\rho_{\odot}$] & $0.16_{-0.04}^{+0.05}$ & $0.16_{-0.03}^{+0.02}$\\
  Surface gravity log $g_{\star}$ [cgs] & $3.95_{-0.09}^{+0.07}$ & $3.96_{-0.05}^{+0.02}$\\
  Mass $M_{\star}$ [$M_{\odot}$] & 1.45 $\pm$ 0.06 & 1.45 $\pm$ 0.05\\
  Radius $R_{\star}$ [$R_{\odot}$] & $2.10_{-0.18}^{+0.24}$ & $2.08_{-0.06}^{+0.12}$\\
  Luminosity $L_{\star}$ [$L_{\odot}$] & $6.8_{-1.3}^{+1.7}$ & $6.8_{-0.8}^{+1.0}$\\
  $u_{1,I+z}$ & 0.18 $\pm$ 0.02 & 0.18 $\pm$ 0.02\\
  $u_{2,I+z}$ & 0.30 $\pm$ 0.01 & 0.30 $\pm$ 0.01\\
  $u_{1,\mathrm{Gunn}-r'}$ & 0.29 $\pm$ 0.02 & 0.29 $\pm$ 0.02\\
  $u_{2,\mathrm{Gunn}-r'}$ & 0.32 $\pm$ 0.01 & 0.32 $\pm$ 0.01\\
  \hline
  \textbf{Deduced planet parameters} & & \\
  \hline
  RV $K$ [$\mathrm{m\:s^{-1}}$] & $53.4_{-6.6}^{+6.8}$ & 52.4 $\pm$ 6.6\\
  Planet/star radius ratio $R_{\mathrm{p}}/R_{\star}$ & 0.084 $\pm$ 0.002 & 0.084 $\pm$ 0.002\\
  $b_{\mathrm{tr}}$ [$R_{\star}$] & $0.24_{-0.16}^{+0.15}$ & 0.23 $\pm$ 0.15\\
  $b_{\mathrm{oc}}$ [$R_{\star}$] & $0.24_{-0.15}^{+0.16}$ & 0.23 $\pm$ 0.15\\
  $T_{\mathrm{oc}}$ [$\mathrm{HJD_{TDB}}$] & $2456472.135_{-0.220}^{+0.146}$ & 2456477.20854 $\pm$ 0.00086\\
  Scaled semi-major axis $a/R_{\star}$ & $6.58_{-0.60}^{+0.56}$ & $6.64_{-0.34}^{+0.17}$\\
  Orbital semi-major axis $a$ [AU] & $0.06432_{-0.00083}^{+0.00088}$ & $0.06431_{-0.00062}^{+0.00065}$\\
  Orbital inclination $i_{\mathrm{p}}$ [deg] & $87.9_{-1.6}^{+1.4}$ & $88.0_{-1.5}^{+1.4}$\\
  Orbital eccentricity $e$ & $0.082_{-0.057}^{+0.084}$, $<$ 0.482 (95 $\%$) & 0 (fixed)\\
  Argument of periastron $\omega$ [deg] & $191_{-79}^{+75}$ & - \\
  Mean density $\rho_{\mathrm{p}}$ [$\rho_{\mathrm{Jup}}$] & 0.11 $\pm$ 0.04 & 0.11 $\pm$ 0.03\\
  Surface gravity log $g_{\mathrm{p}}$ [cgs] & $2.67_{-0.11}^{+0.10}$ & $2.67_{-0.08}^{+0.07}$\\
  Mass $M_{\mathrm{p}}$ [$M_{\mathrm{Jup}}$] & 0.57 $\pm$ 0.08 & 0.56 $\pm$ 0.08\\
  Radius $R_{\mathrm{p}}$ [$R_{\mathrm{Jup}}$] & $1.72_{-0.16}^{+0.21}$ & $1.70_{-0.07}^{+0.13}$\\ 
  Roche limit $a_{\mathrm{R}}$ $\mathrm{[AU]}^{b}$ & $0.02470_{-0.00250}^{+0.00330}$ & $0.02464_{-0.00150}^{+0.00210}$\\
  $a/a_{\mathrm{R}}$ & $2.60_{-0.29}^{+0.28}$ & $2.61_{-0.19}^{+0.16}$\\
  Equilibrium temperature $T_{\mathrm{eq}}$ $\mathrm{[K]}^{c}$ & $1775_{-83}^{+93}$ & $1772_{-45}^{+54}$\\
  Irradiation [erg $\mathrm{s}^{-1} \mathrm{cm}^{-2}$] & $2.3^{+0.9}_{-1.1}\:10^{9}$ & $2.2^{+0.5}_{-0.3}\:10^{9}$\\
  \hline
  \hline
\end{tabular}
\vspace{1cm}
\label{w88table}
\end{table*}

\newpage
\section{Discussion and summary}
\label{discussion}

\begin{figure}[h!]
\caption{Irradiation-radius diagram for the known transiting hot Jupiters with $0.5<M_{\mathrm{p}}<2$ $M_{\mathrm{Jup}}$ and $P<12\:\mathrm{d}$ (data from the NASA Exoplanet Archive). WASP-68 b, WASP-73 b and WASP-88 b are shown in red.}
\centering   
\includegraphics[angle=0, width=0.48\textwidth, height=8.5cm]{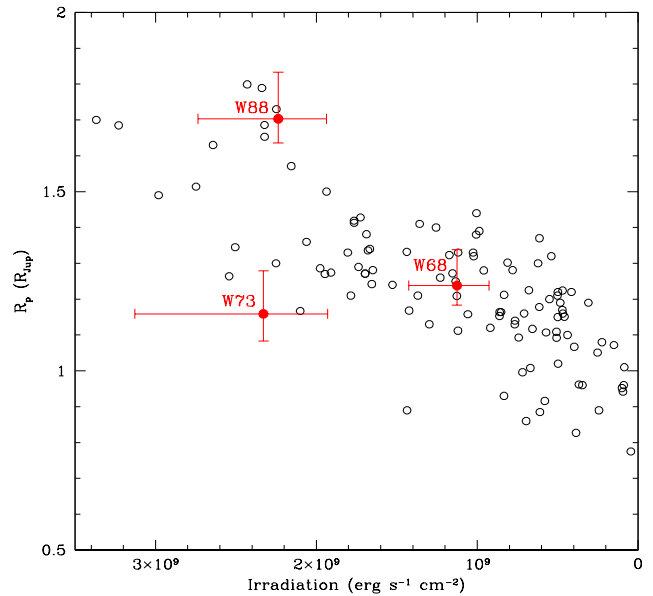}    
\vspace{-0.4cm}       
\label{diagr}
\end{figure}

We presented three newly discovered transiting hot Jupiters from the WASP survey, WASP-68 b, WASP-73 b and \hbox{WASP-88 b}. We derived the parameters of each system from a joint analysis of the CORALIE spectroscopy and the high-precision photometry from TRAPPIST and EulerCam.\\
\indent
All three host stars appear to be significantly evolved (especially WASP-73, see Section \ref{val}) and thus have relatively large radii (1.7-2.2 $R_{\mathrm{\odot}}$). At the time of writing, only about 20 transiting hot Jupiters have been found to orbit such large stars\footnote{\url{http://exoplanet.eu/}}. This small number of detections might be due to the fact that large stellar radii translate into relatively shallow transits for the potential orbiting planets. These transits are therefore more difficult to detect by ground-based transit surveys. Alternatively, this might also be indicative of the tidal destruction of hot Jupiters. Indeed, due to their relatively large masses and small semi-major axes, hot Jupiters are expected to undergo tidal transfers of angular momentum with their host stars (e.g. \citealt{barker}), which should lead in most cases to a slow spiral-in of the planets, until they are finally disrupted at their Roche limits (\citealt{gu}). Although the timescale for orbital decay is quite uncertain and different for each system, it would thus not be surprising to find fewer close-in giant planets around larger and older stars, as there is a higher probability that these planets, if they once existed, have already been tidally disrupted. However, this correlation between the occurrence of hot Jupiters and the \hbox{systems' ages} has not been demonstrated yet.\\
\indent
It is also now common knowledge that tidal interactions tend to realign hot Jupiters' orbits with their host stars' equatorial planes (see e.g. \citealt{barker}). \cite{triaud} demonstrated, using spin/orbit measurements for 22 hot Jupiters around stars with \hbox{masses $\ge$ 1.2  $M_{\mathrm{\odot}}$}, the existence of a correlation between these hot Jupiters' orbital obliquities and their ages, and estimated the typical timescale for a non-coplanar hot Jupiter's orbit to tidally realign to be about \hbox{2.5 Gyr}. Considering the estimated ages of our three systems, we can therefore expect their orbits to have realigned. However, it would be interesting to perform Rossiter-McLaughlin effect observations to confirm this tendency.\\
\indent
Due to the large radii of their host stars, our three planets are exposed to a relatively high irradiation (incident flux \hbox{$>10^{9}$ erg $\mathrm{s}^{-1} \mathrm{cm}^{-2}$}). Several works showed that hot \hbox{Jupiters' radii} correlate well with their irradiating flux (see e.g. \citealt{demory}, \citealt{enoch2} or \citealt{weiss}). Figure \ref{diagr} shows the positions of our three planets in an irradiation-radius diagram for the known transiting planets with $0.5<M_{\mathrm{p}}<2$ $M_{\mathrm{Jup}}$ and $P<12\:\mathrm{d}$. \hbox{WASP-68 b} lies in a well-populated region of the diagram. Its physical dimensions can be considered as rather standard. Indeed, its measured radius of \hbox{$1.24_{-0.06}^{+0.10}$ $R_{\mathrm{Jup}}$} is in perfect agreement with the value of 1.24$\pm$0.03 $R_{\mathrm{Jup}}$ predicted by the equation derived by \cite{weiss} from a sample of 103 transiting planets with a mass between \hbox{150 $M_{\oplus}$} and \hbox{13 $M_{\mathrm{Jup}}$} and relating \hbox{planets' sizes} to their masses and irradiations. For WASP-73 b, the Weiss et al.'s law gives a radius of \hbox{$1.29_{-0.02}^{+0.04}$ $R_{\mathrm{Jup}}$}, which is slightly larger than our measured radius of $1.16_{-0.08}^{+0.12}$ $R_{\mathrm{Jup}}$. This might suggest a possible small enrichment of the planet in heavy elements. Its density of \hbox{$1.20_{-0.30}^{+0.26}$ $\rho_{\mathrm{Jup}}$} is indeed surprisingly high given the important irradiation the planet is exposed to \hbox{($\sim$2.3 $10^{9}$ erg $\mathrm{s}^{-1} \mathrm{cm}^{-2}$)}. However, the errors on its physical parameters are still too high to draw any strong inference on its internal structure.

\indent
Unlike the first two planets, WASP-88 b appears to be a clear outlier, its measured radius of $1.70_{-0.07}^{+0.13}$ $R_{\mathrm{Jup}}$ being significantly higher than the predicted value of 1.35$\pm$0.02 $R_{\mathrm{Jup}}$. With a density of 0.11$\pm$0.03 $\rho_{\mathrm{Jup}}$, WASP-88 b is actually the second least dense transiting planet known to date, tied with Kepler-12 b (\citealt{kep12}), which also has a density around 0.11 $\rho_{\mathrm{Jup}}$. Only WASP-17 b has a lower density ($\rho_{\mathrm{p}}$ = 0.06 $\rho_{\mathrm{Jup}}$, \citealt{W17}, \citealt{W17bis}). WASP-88 b thus joins the handful of planets with super-inflated radii. Its large radius might be explained, at least partially, by the low metallicity ([Fe/H]=-0.08$\pm$0.12) of its star. Indeed, with a mass of 0.56 $M_{\mathrm{Jup}}$, WASP-88 b is actually more Saturn-like than Jupiter-like and \cite{enoch3}, basing on 18 transiting exoplanets with masses below 0.6 $M_{\mathrm{p}}$, found that there is a strong negative correlation between the star metallicity [Fe/H] and $R_{\mathrm{p}}$ for these planets. Keeping in mind that the chemical composition of a planet should be related to the one of its host star as they formed from the same cloud, the fact that the correlation between [Fe/H] and $R_{\mathrm{p}}$ is negligible for more massive planets agrees well with the theoretical planet models of \cite{fortney2} and \cite{baraffe}, which both suggest that the radius of a planet is more sensitive to its composition for low mass planets than it is for more massive ones. The relation (3) of \cite{enoch3} leads to a predicted radius of \hbox{1.51$\pm$0.06 $R_{\mathrm{Jup}}$} for WASP-88 b, which is better than the value of \cite{weiss} but still lower than our measured value. WASP-88 b being the youngest of our three planets (see Section \ref{val}), tidal circularization and dissipation might have occurred recently enough to contribute to the observed inflated radius (see e.g. \citealt{leconte}). Other physical mechanisms might also be at play such as the deposition of kinetic energy into the planet from strong winds driven by the large day/night temperature contrast (\citealt{showman}), reduced heat transport efficiency by layered convection inside the planet (\citealt{chabrier}), or Ohmic heating from currents induced through winds in the planetary atmosphere (\citealt{batygin}).\\



\begin{acknowledgements}

WASP-South is hosted by the South African Astronomical Observatory and we are grateful for their ongoing support and assistance. Funding for WASP comes from consortium universities and from UK's Science and Technology Facilities Council. TRAPPIST is a project funded by the Belgian Fund for Scientific Research (Fonds National de la Recherche Scientifique, F.R.S.-FNRS) under grant FRFC 2.5.594.09.F, with the participation of the Swiss National Science Fundation (SNF). L. Delrez acknowledges the support of the F.R.I.A. fund of the FNRS. M. Gillon and E. Jehin are FNRS Research Associates. A. H. M. J. Triaud received funding from a fellowship provided by the Swiss National Science Foundation under grant number PBGEP2-14559.
\end{acknowledgements}

\vspace{-0.5cm}

\bibliographystyle{aa}
\bibliography{WASP_68_73_88}

\end{document}